\documentclass[twocolumn]{aastex63}
\usepackage{hyperref}
\usepackage{subfigure}
\usepackage{natbib}
\usepackage{xcolor}

\definecolor{edit}{RGB}{178, 34, 34}

\shorttitle{Continuum Lag Investigation}
\shortauthors{Sharp et al.}

\begin{document}

\def\Msun{\hbox{M$_{\odot}$}}
\def\Lsun{\hbox{L$_{\odot}$}}
\def\kms{km~s$^{\rm -1}$}
\def\micron{$\mu$m}
\def\tCO{$^{13}$CO}
\def\nh3{NH$_{3}$}
\def\deg{$^{\circ}$}
\def\arcsec{$^{\prime\prime}$}
\def\arcmin{$^{\prime}$}
\def\Vlsr{\hbox{V$_{LSR}$}}

\newcommand{\CIV}{\hbox{{\rm C}\kern 0.1em{\sc IV}}}
\newcommand{\CIII}{\hbox{{\rm C}\kern 0.1em{\rm [}{\sc III}{\rm ]}}}
\newcommand{\MgII}{\hbox{{\rm Mg}\kern 0.1em{\sc II}}}
\newcommand{\OIII}{\hbox{{\rm [O}\kern 0.1em{\sc iii}{\rm ]}}}
\newcommand{\Hb}{\hbox{{\rm H}$\beta$}}
\newcommand{\Ha}{\hbox{{\rm H}$\alpha$}}
\newcommand{\FeII}{\hbox{{\rm Fe}\kern 0.1em{\sc II}}}
\newcommand{\HeII}{\hbox{{\rm He}\kern 0.1em{\sc II}}}

\title{The Sloan Digital Sky Survey Reverberation Mapping Project: Investigation of Continuum Lag Dependence on Broad-Line Contamination and Quasar Properties}

\correspondingauthor{Hugh W. Sharp}
\email{hugh.sharp@uconn.edu}

\author[0000-0001-9616-1789]{Hugh W. Sharp}
\affiliation{Department of Physics, 196 Auditorium Road, Unit 3046, University of Connecticut, Storrs, CT 06269 USA}

\author[0000-0002-0957-7151]{Y. Homayouni}
\affiliation{Department of Astronomy \& Astrophysics, 525 Davey Lab, The Pennsylvania State University, University Park, PA 16802, USA}
\affiliation{Institute for Gravitation and the Cosmos, The Pennsylvania State University, University Park, PA 16802, USA}

\author[0000-0002-1410-0470]{Jonathan R. Trump}
\affiliation{Department of Physics, 196 Auditorium Road, Unit 3046, University of Connecticut, Storrs, CT 06269 USA}


\author[0000-0002-6404-9562]{Scott F. Anderson}
\affiliation{Astronomy Department, University of Washington, Box 351580, Seattle, WA 98195, USA}

\author[0000-0002-9508-3667]{Roberto J. Assef} \affiliation{Instituto de Estudios Astrofísicos, Facultad de Ingeniería y Ciencias, Universidad Diego Portales, Avenida Ej\'ercito Libertador 441, Santiago, Chile}

\author[0000-0002-0167-2453]{W. N. Brandt}
\affiliation{Department of Astronomy \& Astrophysics, 525 Davey Lab, The Pennsylvania State University, University Park, PA 16802, USA}
\affiliation{Institute for Gravitation and the Cosmos, The Pennsylvania State University, University Park, PA 16802, USA}
\affiliation{Department of Physics, 104 Davey Lab, The Pennsylvania State University, University Park, PA 16802, USA}

\author[0000-0001-9776-9227]{Megan C. Davis}
\affiliation{Department of Physics, 196 Auditorium Road, Unit 3046, University of Connecticut, Storrs, CT 06269 USA}

\author[0000-0001-8032-2971]{Logan B. Fries}
\affiliation{Department of Physics, 196 Auditorium Road, Unit 3046, University of Connecticut, Storrs, CT 06269 USA}

\author[0000-0001-9920-6057]{Catherine J. Grier}
\affiliation{Department of Astronomy, University of Wisconsin-Madison, Madison, WI 53706, USA} 

\author[0000-0002-1763-5825]{Patrick B. Hall}
\affiliation{Department of Physics and Astronomy, York University, 4700 Keele St., Toronto, ON M3J 1P3, Canada}

\author[0000-0003-1728-0304]{Keith Horne}
\affiliation{SUPA Physics and Astronomy, University of St. Andrews, Fife, KY16 9SS, Scotland, UK}

\author[0000-0002-6610-2048]{Anton M. Koekemoer}
\affiliation{Space Telescope Science Institute, 3700 San Martin Dr., Baltimore, MD 21218, USA}

\author[0000-0002-7843-7689]{Mary Loli Martínez-Aldama}
\affiliation{Astronomy Department, Universidad de Concepción, Barrio Universitario S/N, Concepción 4030000, Chile}

\author[0000-0003-4010-5336]{David M. Menezes}
\affiliation{Department of Physics, 196 Auditorium Road, Unit 3046, University of Connecticut, Storrs, CT 06269 USA}

\author[0000-0002-0033-5041]{Theodore Pena}
\affiliation{Department of Astronomy, University of Wisconsin-Madison, Madison, WI 53706, USA}

\author[0000-0001-5231-2645]{C. Ricci}
\affiliation{Instituto de Estudios Astrofísicos, Facultad de Ingeniería y Ciencias, Universidad Diego Portales, Avenida Ej\'ercito Libertador 441, Santiago, Chile}
\affiliation{Kavli Institute for Astronomy and Astrophysics, Peking University, Beijing 100871, China}

\author[0000-0001-7240-7449]{Donald P.\ Schneider} 
\affiliation{Department of Astronomy \& Astrophysics, 525 Davey Lab, The Pennsylvania State University, University Park, PA 16802, USA}
\affiliation{Institute for Gravitation and the Cosmos, The Pennsylvania State University, University Park, PA 16802, USA}

\author[0000-0003-1659-7035]{Yue Shen}
\affiliation{Department of Astronomy, University of Illinois at Urbana-Champaign, Urbana, IL 61801, USA}
\affiliation{National Center for Supercomputing Applications, University of Illinois at Urbana-Champaign, Urbana, IL 61801, USA}

\author[0000-0002-3683-7297]{Benny Trakhtenbrot}
\affiliation{School of Physics and Astronomy, Tel Aviv University, Tel Aviv 69978, Israel}


\begin{abstract}

This work studies the relationship between accretion-disk size and quasar properties, using a sample of 95 quasars from the SDSS-RM project with measured lags between the $g$ and $i$ photometric bands. Our sample includes disk lags that are both longer and shorter than predicted by the \citet{SS73} model, requiring explanations which satisfy both cases. Although our quasars each have one lag measurement, we explore the wavelength-dependent effects of diffuse broad line region (BLR) contamination through our sample's broad redshift range, $0.1<z<1.2$. We do not find significant evidence of variable diffuse \FeII\ and Balmer nebular emission in the root-mean-square (RMS) spectra, nor from Anderson-Darling tests of quasars in redshift ranges with and without diffuse nebular emission falling in the observed-frame filters. Contrary to previous work, we do not detect a significant correlation between measured continuum and BLR lags in our luminous quasar sample, similarly suggesting that our continuum lags are not dominated by diffuse nebular emission. Similar to other studies, we find that quasars with larger-than-expected continuum lags have lower 3000~\AA\ luminosity, and we additionally find longer continuum lags with lower X-ray luminosity and black hole mass. Our lack of evidence for diffuse BLR contribution to the lags indicates that the anti-correlation between continuum lag and luminosity is not likely to be due to the Baldwin effect. Instead, these anti-correlations favor models in which the continuum lag increases in lower-luminosity AGN, including scenarios featuring magnetic coupling between the accretion disk and X-ray corona, and/or ripples or rims in the disk.

\end{abstract}

\section{Introduction}
\label{sec:intro}
Active galactic nuclei (AGN) are the most luminous persistent sources of radiation in our universe, and are characterized by non-stellar spectra that are driven by an accreting disk of matter falling toward a super massive black hole (SMBH) \citep{Bell1969}. Observations demonstrate that all massive galaxies have a central SMBH (\textit{e.g.}, \citealp{Magorrian1998, Kormendy2013}), indicating that AGN, as rapidly-growing SMBHs, are important to the field of galaxy evolution as a whole \citep{DiMatteo2003, DiMatteo2005, Hopkins2005a, Hopkins2005b}. The majority of SMBH growth is governed by rapid accretion (\textit{e.g.}, \citealp{Soltan1982}), so understanding the detailed geometry and emission profile of these disks is critically important for SMBH buildup and its connection to galaxy evolution. 

In general, the physical components that comprise the innermost regions of AGN are not spatially resolvable (for exceptions, see \citealp{EHT2019, EHT2022, GRAVITY2018}). Thus measurement of AGN scale and structure most commonly relies on Reverberation Mapping (RM; \textit{e.g.},\citealp{Blandford1982, peterson2004, Cackett2021}), a method utilizing time-domain monitoring to substitute temporal resolution for spatial resolution. The method relies on the characteristic variability of a quasar's central emission being re-emitted (``reverberated'') by more distant material, delayed (or ``lagged'') by the light-crossing time ($\tau = R/c$) of the system \citep{Cackett2007, Cackett2021}. The natural temperature gradient of the system causes the peak emission of hotter, more central regions, to occur at shorter wavelengths, while regions at larger radii have peak emission at longer wavelengths (\textit{e.g.}, \citealp{Shappee2014, McHardy2014, Sergeev2005, Collier1998}). By measuring the time delays between variability features in light curves observed in various wavelengths, we can recover information on the scale and structure of various AGN components, characterized by the relative locations of their observed emission. The RM method was first implemented to measure the size of the broad-line region (BLR), but can be used to measure other parts of the AGN inner environment, including the X-ray corona, accretion disk, and dusty torus \citep[][and references therein]{Cackett2021}. In particular, the accretion disk structure is probed by utilizing continuum RM, where the lag between variability features is measured between multiple UV and optical continuum light curves, driven by (unobserved) central X-ray ionization \citep{Cackett2007}.


For an idealized geometrically-thin, optically-thick, steady-state accretion disk, the \citet[][e.g., SS73]{SS73} model describes the relationship of disk size with black hole mass, accretion rate, and emission wavelength:


\begin{equation}
    \label{eq:1}
    \tau \propto (M_{\rm BH}\dot{M})^{1/3} \lambda^{4/3}
\end{equation}
where $\tau=R/c$ is the observed lag and $\lambda \propto 1/T$ for blackbody peak emission, where T is the temperature of the disk. For accretion-disk lag measurements in large continuum RM surveys, the above equation is often fit by the function $\tau = \tau_0[(\lambda/\lambda_0)^\beta - 1]$, where $\tau_0$ is a normalization factor, $\lambda_0$ is a reference wavelength, and $\beta$ represents the wavelength exponent in equation~(\ref{eq:1}). In the case of local, single-object studies \citep{McHardy2014, Edelson2015, Fausnaugh2016, Edelson2017, McHardy2018}, the wavelength dependence of the disk lag appears to agree with that of SS73~(Equation \ref{eq:1}), where $\beta\sim4/3$. The normalization factor indicative of scale, $\tau_0$, on the other hand, has generally been reported to be $\sim$2-3 times larger than anticipated. 

UV/optical observations of continuum RM require a high temporal resolution ($\sim$1~day) due to the small relative distance between wavelength regions. This requirement has limited the application of large surveys to study accretion disk size over a large sample until very recently. Large surveys like Pan-STARRS \citep{Jiang2017}, DES \citep{Mudd2018, Yu2020}, and SDSS \citep{Homayouni2019} have found that $\beta \sim 4/3$; thus the wavelength dependence of the lag is in agreement with Equation~\ref{eq:1}.
The normalization factor $\tau_0$, conversely has had differing results from the SS73 prediction among these surveys.
Of these larger studies mentioned, \citet{Jiang2017} and \citet{Mudd2018} find $\tau_0$ to be $\sim$2-3 times larger than anticipated, similar to single-object studies. \citet{Yu2020} and \citet{Homayouni2019} argue the average disk lag of their surveys agree with SS73, but with large scatter in individual disk lags that significantly exceeds the observational uncertainties. \citet{Homayouni2019} also discusses that the larger lags reported by other surveys may be due to biases towards large lags in cadence-limited observations.

In addition, a few accretion disk scales have been observed via gravitational microlensing of quasars, a different process entirely from continuum RM. In cases such as \citet{Morgan2010}, the optical emitting region representing the disk is larger than expected, similar to the general findings of continuum RM. These measurements could however suffer from  observational biases due to a combination of the inclination of the disk relative to the line of sight and the differential magnification of the temperature fluctuations, producing overestimated disk sizes \citep{Tie2018}.

These larger than anticipated disk sizes have many possible explanations. The SS73 model may simply be over idealized; for example, it does not properly describe basic features of quasars such as variability \citep{Dexter2010}. Many studies include different disk structure and/or radiative-transfer effects that produce larger disk sizes than predicted by the SS73 model. \citet{Dexter2010} suggest that local fluctuations in the accretion disk produce a larger overall disk size. \citet{Hall2018} also show that with a sufficiently low accretion disk atmospheric density, scattering in the atmosphere can produce non-blackbody emergent spectrum, also resulting in larger disk lags. Additionally \citet{Mummery2020} suggest that larger disk lags can occur in disks dominated by tidal disruption events (TDEs). \citet{Starkey2023} have also demonstrated that invoking a disk geometry with a steep rim or rippled structures will result in increased irradiated luminosity, and thus temperature, producing reverberation lags that are larger than the simple SS73 thin disk picture.

Different emission reprocessing models have also been shown to produce larger continuum lags. \citet{Kammoun2019} and \citet{Kammoun2021} study disk-reprocessing models with general relativistic ray-tracing and find that a larger X-ray corona height tends to yield systematically larger lags. \citet{Sun2020} introduce the corona-heated accretion disk reprocessing model (CHAR model), in which the corona and the accretion disk are coupled via a magnetic field. Energy transfer by magnetic heating adds an additional time delay for disk reprocessing that is related to the thermal timescale ($\tau_{\rm TH}$) in addition to the light crossing time, increasing the lags for lower-mass black holes in particular. In addition, there is potential for nebular continuum emission to affect the measured disk lag. As discussed by \citet{Cackett2018}, photometric observations can be contaminated by diffuse emission from the broad-line region (BLR), resulting in increased lag localized to these diffuse emitting regions. \citet{Chelouche2019} and \citet{Cackett2022} argue that all UV/optical photometric filters may be contaminated by substantial diffuse continuum emission from the boundary of the outer accretion disk and BLR.

\begin{figure}[ht!]
\centering
    \includegraphics[width=\columnwidth]{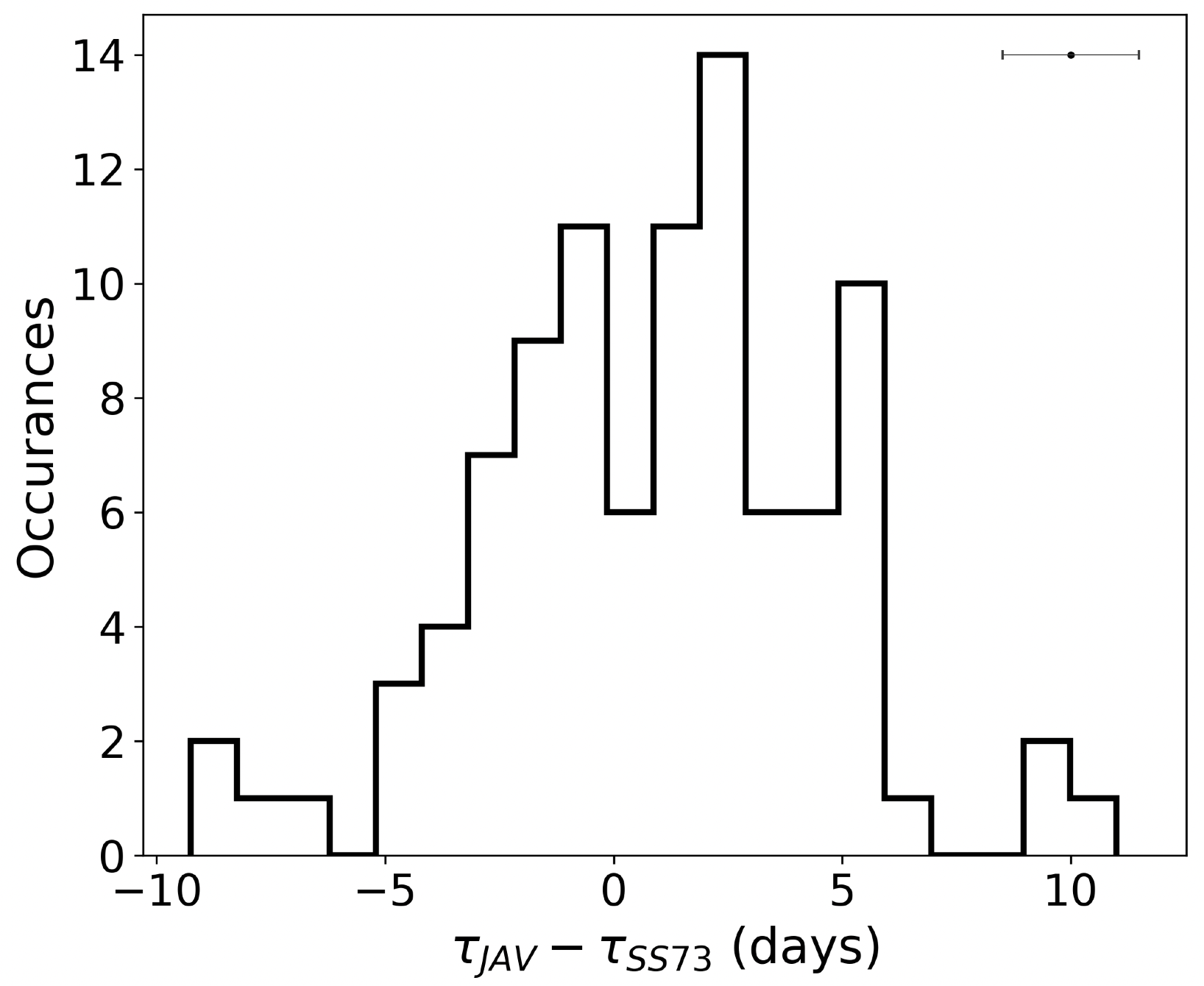}
\caption{The distribution of the offset between the observed disk lags obtained using \texttt{JAVELIN}, and the disk lag anticipated by SS73 using equations~(\ref{eq:2}) and~(\ref{eq:3}). Only 36\% of the sample fall within 1$\sigma$ of the SS73 prediction, and addtionally have an excess scatter of $\sigma\tau / \tau = 1.35$, implying there is intrinsic scatter in disk size within this quasar sample.
}\label{fig1}
\end{figure}

To better understand the cause behind these larger disk sizes, we study the detailed quasar properties for the 95 objects with measured disk lags from \citet{Homayouni2019}. While the \textit{average} lag of this sample was found to be consistent with SS73, the lags have a much broader distribution than expected given the measurement uncertainties, with an excess scatter of $\sigma\tau / \tau = 1.35$ (see Figure~\ref{fig1}). These quasars were monitored as a part of the Sloan Digital Sky Survey Reverberation project (SDSS-RM; \citealp{Shen2015}), utilizing both broad-line RM \citep{Grier2017} to obtain black hole mass ($M_{\rm BH}$) measurements, and continuum RM in the $g$ and $i$-bands to measure accretion-disk scale and structure \citep{Homayouni2019}. In addition, these 95 SDSS targets cover a large range of quasar properties, such as black hole mass, luminosity ($L$), Eddington ratio ($\lambda_{\rm Edd} = L/L_{\rm Edd}$) each spanning $\sim$3 magnitudes, among others provided by \cite{Shen2019}, over the redshift range of $0.1<z<1.2$. Studying these disk lags and their deviation from SS73 amongst such a broad quasar demographic can expose systematic trends and reveal necessary accretion physics missing from the model.

In Section~\ref{sec2}, we discuss the sample of 95 quasars used in \citet{Homayouni2019}, including observations, measurement of disk lags, and construction of root-mean-square (RMS) spectra. In Section~\ref{sec:3} we discuss patterns and correlations (or the lack thereof) that arise when comparing the observed continuum lags to various quasar properties. We conclude in section~\ref{sec:4} with a discussion of the implications of our observations for models of accretion-disk structure and emission reprocessing.

\section{Data}
\label{sec2}
This study includes 95 SDSS-RM quasars with continuum lags measured by \citet{Homayouni2019} in order to better understand how accretion disk structure depends on quasar properties. The SDSS-RM project \citep{Shen2015} monitored a total of 849 quasars in a 7 deg$^2$ field
. The sample spans a broad range of redshift, black hole mass, luminosity, and other quasar properties \citep{Shen2019}, making it a useful sample for studying the diversity of quasar accretion-disk structure. A subset (44) of the black hole masses are measured from broad-line reverberation mapping by \citet{Grier2017}, while the remainder are estimated from single-epoch scaling relations (as described in \citealt{Shen2019}). We additionally use X-ray luminosities from XMM-Newton imaging of the field \citep{Liu2020}.

\subsection{Observations and light curves}

We use continuum lags and RMS spectra measured from the first year (2014) of SDSS-RM observations. This includes 32 epochs of spectroscopic monitoring from the SDSS/BOSS instrument \citep{Dawson2013}, from which synthetic $g$ and $i$ photometry were extracted using the SDSS filter response functions \citep{Fukugita1996}. The light curves include an additional 63 epochs of photometric monitoring from the Bok~2.3m and CFHT~3.6m telescopes \citep{Kinemuchi2020}. Observations from the three observatories (SDSS, Bok, and CFHT) were inter-calibrated and combined using the \texttt{CREAM} software \citep{Starkey2016}.

\subsection{Disk Lag Measurements}
\label{sec:2.2}

\cite{Homayouni2019} measure the lags between observed-frame $g$ and $i$-band light curves through two methods, an Interpolated Cross Correlation Function (\texttt{ICCF}; \citealp{Gaskell1986, Gaskell1987, White1994, peterson_2004}) and \texttt{JAVELIN} \citep{Zu2011}. \texttt{ICCF} uses simple linear interpolation to reconstruct the lightcurve during monitoring gaps in the observations, and calculates the Pearson coefficient $r$ between them across a range of allowed $\tau$, thus producing a cross correlation function (CCF). \texttt{JAVELIN} more robustly accounts for monitoring gaps by fitting the bluest lightcurve with a damped random walk (DRW) model. Observations have shown that a DRW model is appropriate in describing the variability of a quasar lightcurve on timescales relevant for RM studies \citep{Kelly2009, Kozlowski2010, MacLeod2010}. \texttt{JAVELIN} uses the constructed DRW model of the blue lightcurve and performs a parameterized fit to the responding lightcurve, assuming that the responding lightcurve is a shifted, smoothed, and scaled version of the model, thus recovering $\tau$. 

Each of these methods use Monte Carlo techniques to estimate the lag uncertainty, which are discussed in further detail in \cite{Homayouni2019}. Lag estimates from both methods were found to be consistent in our sample, whereas the lag uncertainties produced from \texttt{ICCF} seem to be overestimated in comparison to those produced by \texttt{JAVELIN}. This is consistent with the results of \citet{Yu2020unc}, which demonstrate that the JAVELIN uncertainties are more accurate and the ICCF uncertainties are overestimated for light curves with the characteristics of those from SDSS-RM. We thus use the \texttt{JAVELIN} lags and uncertainties of the 95 SDSS-RM quasars to explore their diversity in accretion disk size. These lags are considered ``well-defined'' amongst a larger sample of 222 targets, passing three selective criteria presented in \cite{Homayouni2019} to ensure the lags were not subject to measurement bias. These criteria filtered out targets which exhibited insignificant cross-correlation, ambiguous lag detection resulting from bimodal probability distributions, and significant broad-line contribution from \CIV, \MgII, \Hb, and \Ha~ emission lines, with specific quantitative thresholds defined in \cite{Homayouni2019}.

\subsection{Spectroscopic Data and PrepSpec}

All these targets have spectra collected throughout the duration of SDSS-RM observations, each with 90 epochs from 2014 to 2020. We use these data to probe quasar properties associated with emission lines. \texttt{PrepSpec}, described in \citet{Shen2016}, was used to create both time-averaged and RMS spectra in the optical/UV, giving the average and variability amplitude of the spectra over those 90 epochs. \texttt{PrepSpec} decomposes time-resolved spectra into a model considering wavelength and time-dependent components for the continuum and broad and narrow emission lines, creating a model of the mean quasar spectra. \texttt{PrepSpec} also creates model root-mean-squared (RMS) spectra that are a measure of variability of the modeled components over our 90 epochs observed for each of our quasars. As will be discussed further in section \ref{sec:3.2}, we utilize the modeled RMS spectra to investigate variability of the diffuse nebular continuum as a potential bias in the lag measurements of our quasars.

\section{Results}
\label{sec:3}

\subsection{Comparing \texorpdfstring{$\tau_{\rm jav}$ to $\tau_{\rm SS73}$}{tjav to tSS73}}

We compare the measured lags from \citet{Homayouni2019} with the lags predicted by the SS73 model. Observing in the $g$ and $i$-band, the disk lag predicted by the SS73 model is as follows:

\begin{equation}
    \label{eq:2}
    \tau_{\rm SS73} = \frac{\tau_0}{(1+z)^{1/3}} \left[ \left(\frac{\lambda_i}{\lambda_0}\right)^{4/3} - \left(\frac{\lambda_g}{\lambda_0}\right)^{4/3} \right]
\end{equation}

Here \cite{Homayouni2019} use a normalized wavelength of $\lambda_0 = \lambda/9000$ \AA\, and produce a disk normalization, $\tau_0$, representative of a realistic disk size. The analytic form of $\tau_0$ is given as:

\begin{equation}
    \label{eq:3}
    \tau_0 = \frac{1}{c}\left(\frac{45 G}{16 \pi^6 h_p c^2}\right)^{1/3} X^{4/3} (\lambda_0)^{4/3} \left(\frac{C_{\rm Bol}}{\eta c^2}\right)^{1/3} M^{1/3}_{\rm BH} \lambda L_{\lambda 3000}^{1/3}
\end{equation}


Here $C_{\rm Bol} = 5.15$ is the bolometric luminosity correction \citep{Richards2006}, and  $\eta=0.1$ was chosen as the radiative efficiency. The factor $X$ accounts for the wavelength range of blackbody emission at a given temperature, originating from any given accretion disk radii, where $X = \frac{hc}{\lambda kT(\lambda)}$.  We adopt the value $X = 2.49$ which \citet{Fausnaugh2016} derive from a flux-weighted mean radius.


As noted by \citet{Homayouni2019}, the \textit{average} disk lag of the sample is consistent with $\tau_{\rm SS73}$, but with a large scatter. This scatter is greater than the observational uncertainty, with only 36\% of the sample of 95 quasars falling within 1$\sigma$ of the model lags as highlighted in Figure~\ref{fig1}. As \texttt{JAVELIN} has been thoroughly tested to produce accurate uncertainties \citep{Yu2020},  the broad distribution of measured lags indicates genuine excess scatter compared to the SS73 model expectation. We adopted two variables to quantify the observed SS73 model deviations, the disk lag offset ($\tau_{\rm jav} - \tau_{\rm SS73}$) and disk lag ratio ($\tau_{\rm jav} / \tau_{\rm SS73}$), when testing for correlations with quasar properties.

\subsection{Diffuse Contamination}
\label{sec:3.2}

\begin{figure}[ht!]
\centering
    \includegraphics[width=\columnwidth]{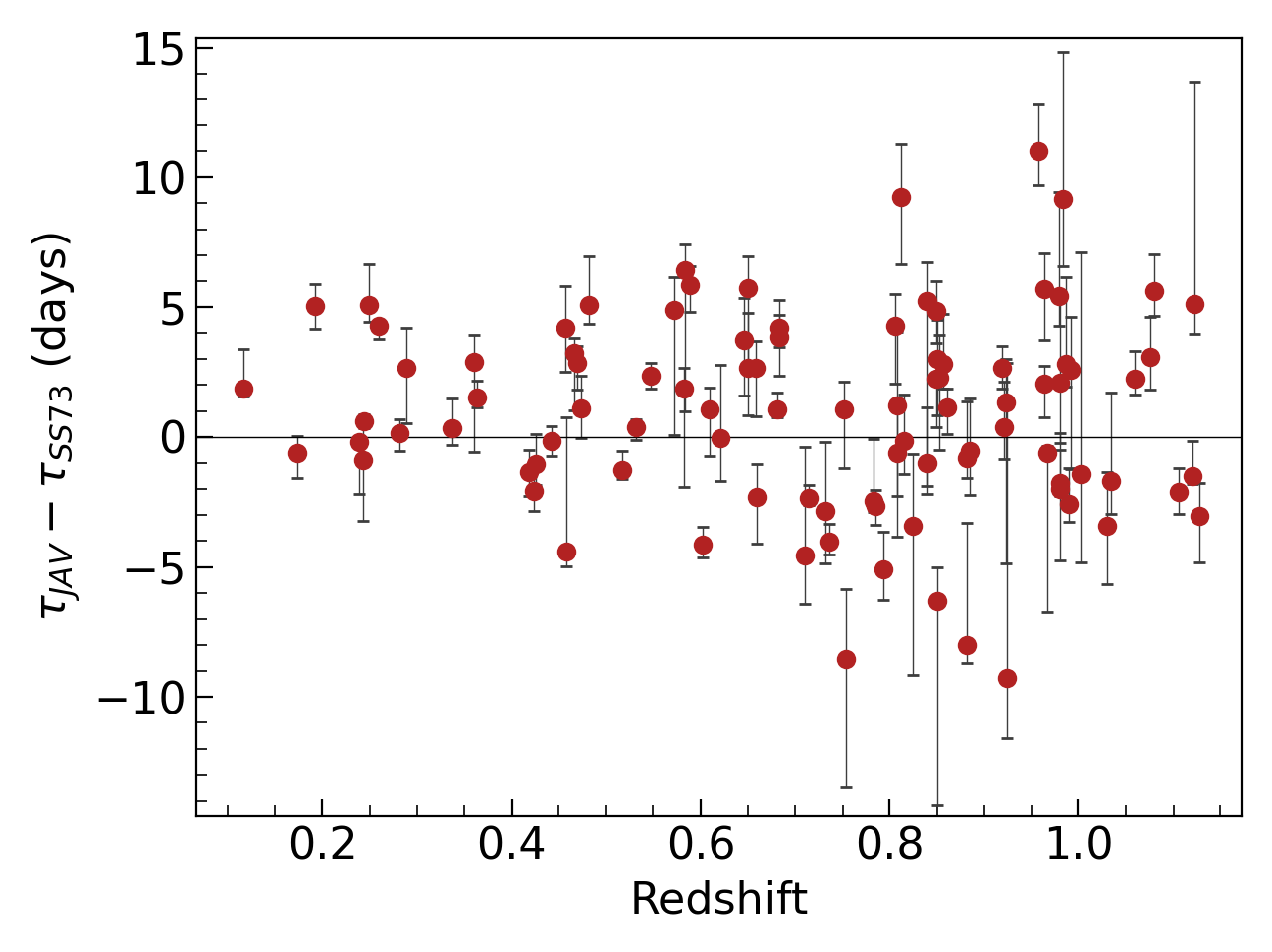}
\caption{The sample of 95 quasars with their disk lag offset ($\tau_{\rm jav}-\tau_{\rm SS73}$) plotted versus redshift. The sample's disk lag offset noticeably increases in scatter within $0.8<z<1.0$. At this redshift range, the $g$ and $i$ photometry bands reside in regions of diffuse Balmer and \FeII\ emission respectively, which may be contributing to the variability in our light curves.}
\label{fig2}
\end{figure}

When comparing the disk lag offset to redshift as seen in Figure~\ref{fig2}, by visual inspection, our quasars seem to exhibit a larger scatter in disk lag offset towards higher $z$, specifically around the range of $0.8<z<1.0$ (henceforth referred to as $z_{\rm con}$). This redshift range is notable because it corresponds to the regime in which our photometry may be contaminated by diffuse \FeII\ emission in the \textit{g}-band, and diffuse Balmer emission in the \textit{i}-band. Diffuse emission from gas in the more distant BLR can cause longer than anticipated lags contributing to the observed $i$-band emission, as well as shorter lags if present in the $g$-band \citep{Netzer2022}. Evidence for longer lags due to diffuse Balmer emission has been found previously in single-target, intensive campaigns with multi-band lag analysis \citep{Edelson2019, Vincentelli2022, Santisteban2020}. For example \cite{Cackett2018} create a ``lag spectrum'' that comprises the measured disk lag as a function of wavelength for NGC 4593, finding a broad excess in the measured lag leading up to the Balmer jump (3640 \AA). \citet{Chelouche2019}, \citet{Cackett2022}, and \citet{Netzer2022} present further evidence that diffuse BLR emission may additionally affect multiple photometric bands in addition to the strong effect from the Balmer continuum.

\begin{figure}[ht!]
\centering
    \includegraphics[width=\columnwidth]{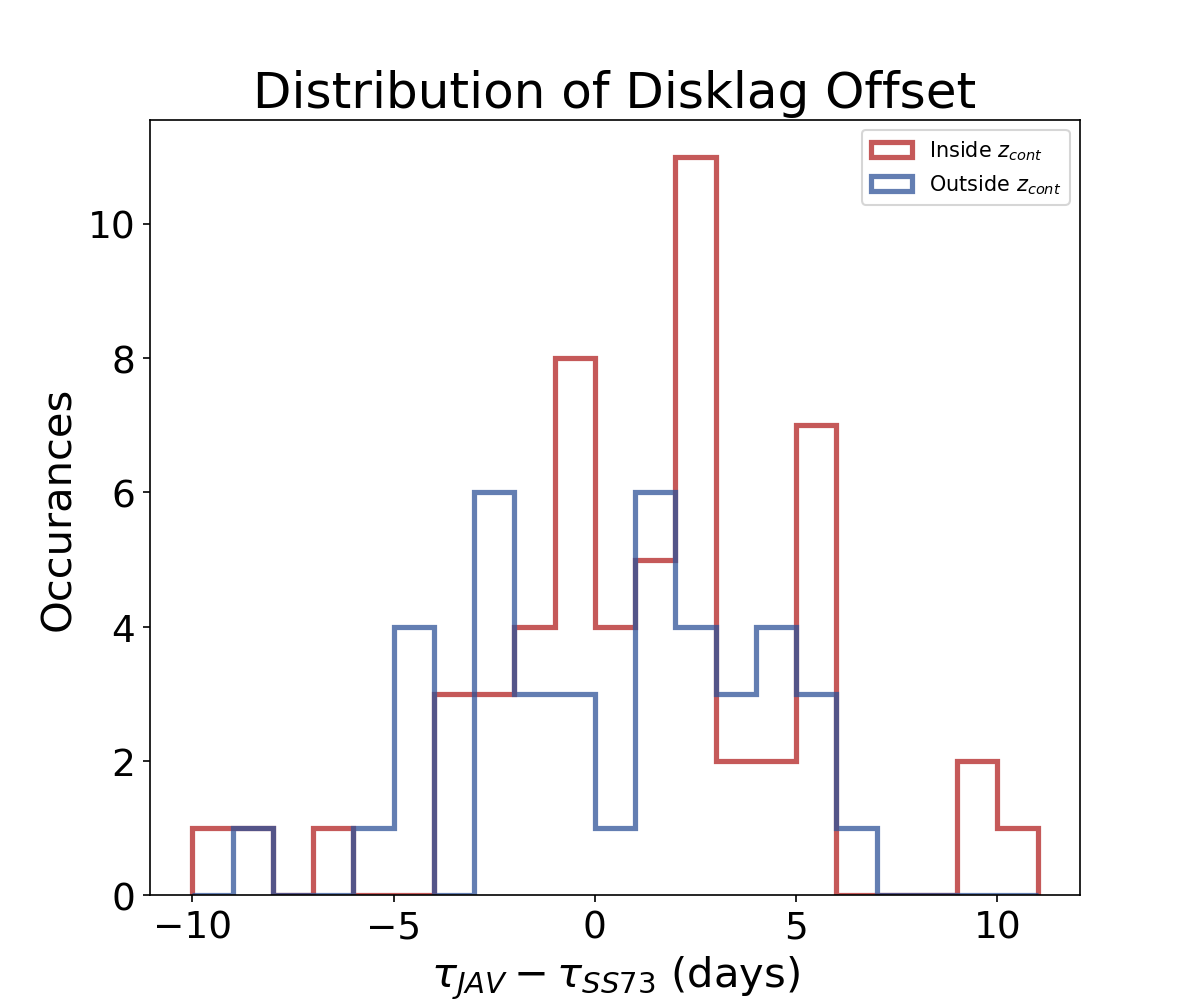}
\caption{Disk lag offset distribution shown in Figure~(\ref{fig1}) split into a distribution of targets within $z_{\rm con}$ (in red) and outside $z_{\rm con}$ (in blue). Our k-sample Anderson-Darling test was unable to reject the null-hypothesis that this distributions are drawn from the same population.}\label{fig3}
\end{figure}

We performed a statistical analysis to determine whether distribution of disk lags were significantly different between the population of quasars inside and outside $z_{\rm con}$. Figure~\ref{fig3} splits the disk lag distribution given in Figure~\ref{fig1} into these regions in redshift. If there is no redshift dependence on disk lag offset, both of these particular quasar distributions should be drawn from a similar parent distribution of disk lag offset. We used a k-sample Anderson-Darling (AD) test to determine whether the distribution of disk lag offset inside and outside $z_{\rm con}$ can be statistically drawn from the same parent population. The k-sample AD tests the null hypothesis that $k$-samples are drawn from the same population without having to specify the distribution function of a parent population, as detailed in \citet{ADTest}. Using the \texttt{scipy.stats} \citep{SciPy} implementation of the k-sample AD test, we find a \textit{p}-value $> 0.05$. This indicates that we cannot confidently reject that the disk lag offset distribution from our 95 quasar sample differs in parent distribution based on redshift. The Anderson-Darling test is inconclusive, we next study the variable spectra of our quasars to search for direct evidence for such contamination.

We also quantified the variability of the diffuse Balmer and Fe~II emission by measuring their equivalent widths (EW) from the RMS spectra. Quasars with significantly variable diffuse BLR emission that biases the measured $g-i$ lag should have a larger contribution to their RMS spectra. Thus measuring these EWs in the RMS spectra should indicate whether these diffuse emission features have substantial variability in comparison to the continuum. We made EW measurements in rest-frame wavelength ranges 3500-4000 \AA\ and 2250-2650 \AA\ to probe for diffuse Balmer and \FeII\, respectively. To ensure complete coverage of the wavelength range for nebular continuum emission (and reliable continuum estimates around the diffuse emission), we restricted the diffuse Balmer sample to $z>0.3$ and the diffuse \FeII\ sample to $z>0.6$, measuring EWs for 84 and 55, respectively,  quasars out of our sample's total of 95. The RMS continuum fit by \texttt{PrepSpec} was normalized to RMS flux around 3000~\AA\ in the rest-frame to avoid prominent emission lines. Figure~\ref{fig5} presents two examples RMS spectra that have significant diffuse continuum emission. We estimate EW uncertainties using a Monte Carlo method, adding random Gaussian noise associated with the uncertainties of the RMS spectra for each pixel in wavelength. 

\begin{figure}[ht!]
\centering
    \includegraphics[width=\columnwidth]{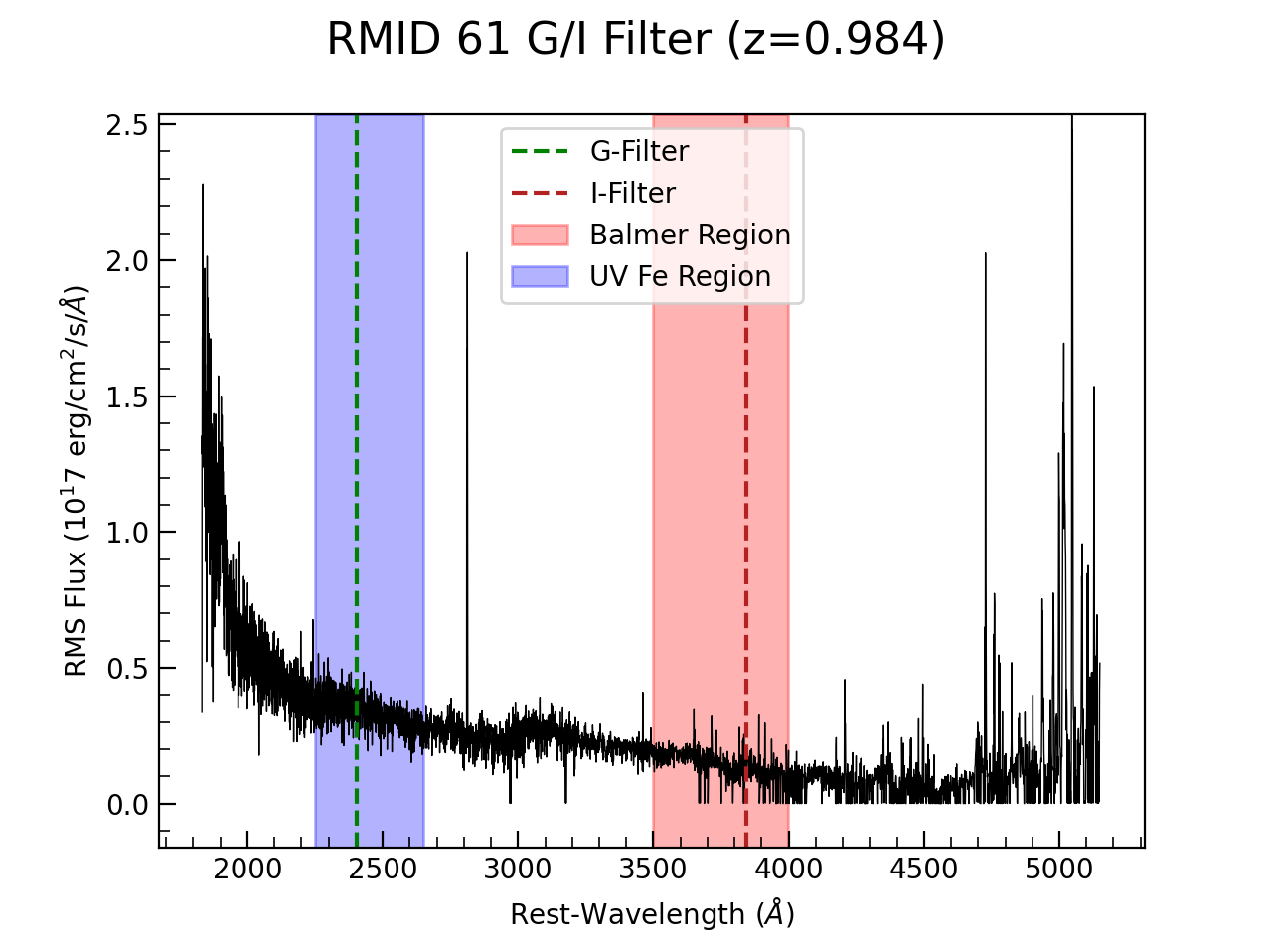}
\caption{An example RMS spectrum of RMID~61, a quasar within $z_{\rm con}$. The blue and red shaded regions represent the wavelength ranges of diffuse \FeII\ and Balmer emission, while the dashed green and red lines represent the central wavelengths of the $g$ and $i$ photometric filters for this quasar As shown here there is possible contamination in RMID~61 in both the $g$ and $i$-bands as they fall within the regions of diffuse UV Fe and Balmer emission, respectively.}\label{fig4}
\end{figure}

Figure~\ref{fig6} reveals our quasar sample spans a large range of diffuse \FeII\ and Balmer EW measurements in the restframe. When fit via \texttt{linmix}'s implementation of linear regression \citep{Kelly2007}, no correlation was detected between either diffuse \FeII\ and Balmer RMS EWs and the disk lag offset of our sample. Even at the extrema of $\tau_{\rm jav} - \tau_{\rm SS73}$, these quasars do not appear to have the significantly larger diffuse Balmer EWs for the longer lags and larger diffuse Fe~II EWs for shorter lags that we would expect if diffuse BLR emission contributes to their lag measurements. Even a visual inspection suggests few of these quasar spectra have diffuse features apparent in the RMS spectra as shown in Figure~\ref{fig5}, and those displayed do not have particularly extreme deviations from their predicted SS73 result. This behaviour is similar to NGC 4593 covered in \cite{Cackett2018}, where the excess lag in the 3000-4000 \AA\ regime implies diffuse Balmer contamination, but there does not appear to be any significant increase in the RMS spectra of NGC 4593 in the 3000-4000 \AA\ range.

\begin{figure}[ht!]
\centering
    \includegraphics[width=\columnwidth]{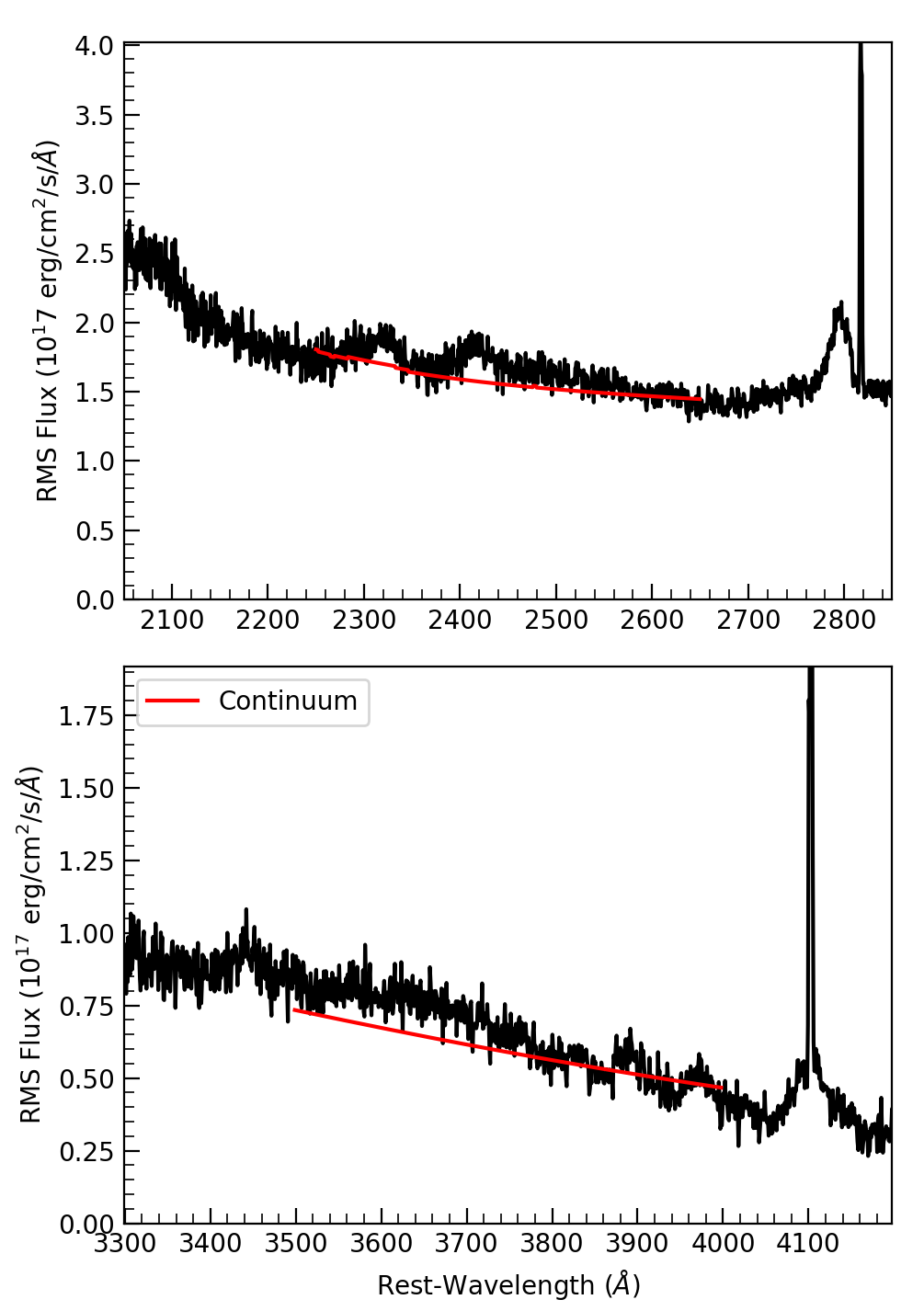}
\caption{Example of targets containing noticeable diffuse contamination in their RMS spectrum (black) as identified from a continuum fit (red lines). In the top plot, RMID 428 has excess variable emission in the 2250-2650 \AA\ region consistent with diffuse Fe UV emission. In the bottom plot, RMID 160 has excess variable emission compared to the continuum over $\sim$3500-3700 \AA\ that is consistent with diffuse Balmer emission.}\label{fig5}
\end{figure}

\begin{figure*}
\centering
    \includegraphics[width=\textwidth]{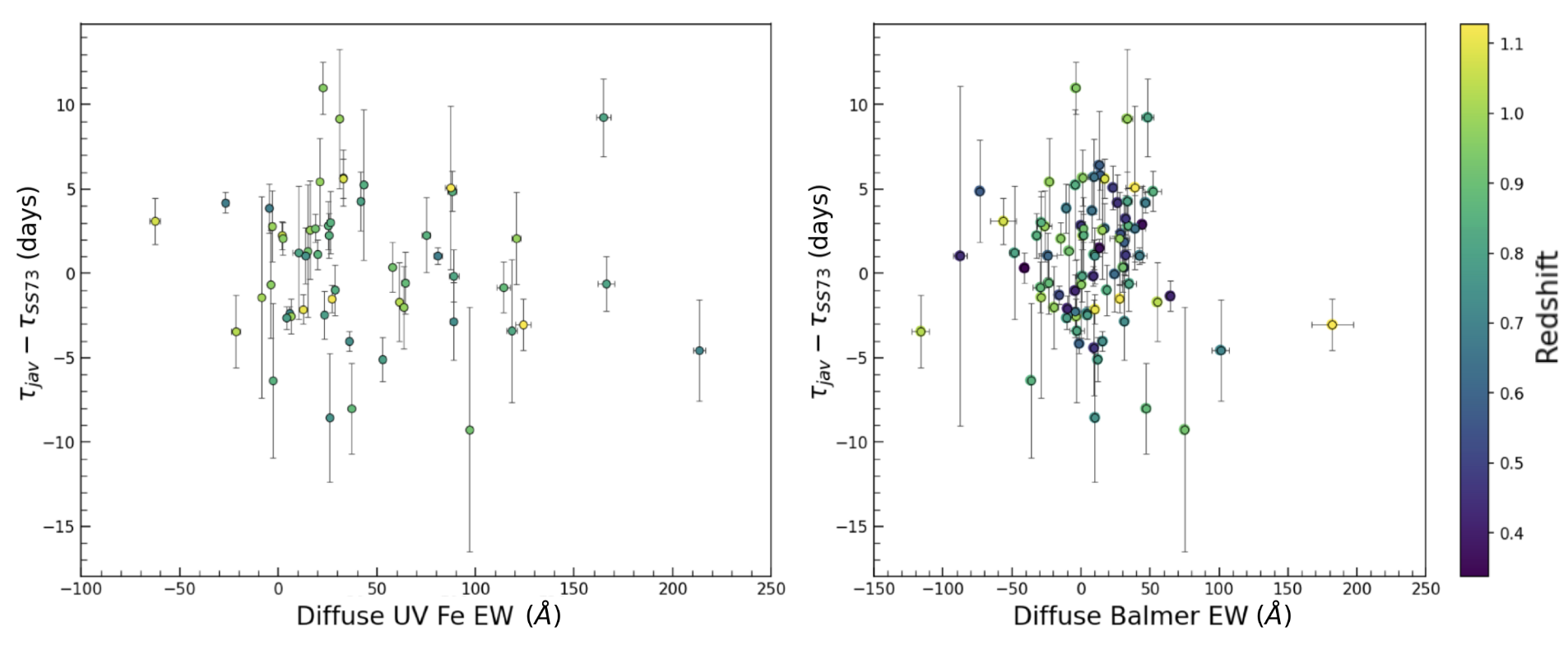}
\caption{The measured restframe EWs of the Fe~II (2250-2650 \AA; \textit{left}) and Balmer continuum (3500-4000 \AA; \textit{right}) in the RMS spectra, plotted against disk lag offset ($\tau_{\rm jav} - \tau_{\rm SS73}$) and colored by redshift. The points represent the mean of the distribution of EWs measured in the Monte Carlo described in section~\ref{sec:3.2}, while the error bars represent the standard deviation. There are 55 quasars with diffuse Fe~II measurements, and 84 quasars with diffuse Balmer EW measurements, due to limited coverage of their respective regions in our RMS spectra based on redshift. When fitting using \texttt{LinMix} as a Bayesian approach to linear regression, there was no correlation between disk lag offset and either of these diffuse EW measurements.} \label{fig6}
\end{figure*}

Lastly, \citet{Wang2023} propose that if continuum lag measurements are dominated by diffuse BLR emission, there should be a tight correlation between continuum and BLR lags. Continuum lags dominated by diffuse BLR emission would further imply a $R_{\rm cont}-L$ relation that is analogous to the $R_{\rm BLR}-L$ correlation, notably used to estimate the bulk of black hole mass growth over cosmic time \citep[e.g.][]{Vestergaard2009}. A relationship between continuum lags and luminosity would be far less observationally demanding than the time-domain spectroscopy required for BLR lag measurements.
\citet{Wang2023} report a correlation between $R_{\rm BLR}-R_{5100}$, which they use to further imply a $R_{5100}-L$ relation, where $R_{5100}$ would be the continuum size at rest-frame 5100 \AA.  
To investigate a similar correlation in the SDSS-RM sample, we use the subset of 30 quasars that both feature well-measured disk sizes from \citet{Homayouni2019} and reliable \Hb\ BLR size measurements from \citet{Grier2017}. We then convert the observed-frame $g-i$ lag to the rest-frame time delay between the inner accretion disk to the rest-frame 5100 \AA\ emitting region [$\tau_{5100} = \tau_0 (\lambda_0 = 5100\text{ \AA})$].
Similar to \citet{Wang2023}, we assume a wavelength dependence of $\beta=4/3$ and convert the $g-i$ lag measurements as shown below, following a similar form of Equation~\ref{eq:2}.

\begin{equation}
\label{eq:4}
\tau_{5100} = \tau_{jav} (1+z)^{1/3} \left[ \left(\frac{\lambda_i}{5100\text{\AA}}\right)^{4/3} - \left(\frac{\lambda_g}{5100\text{\AA}}\right)^{4/3} \right]^{-1}
\end{equation}

\begin{figure}
\centering
\includegraphics[width=\columnwidth]{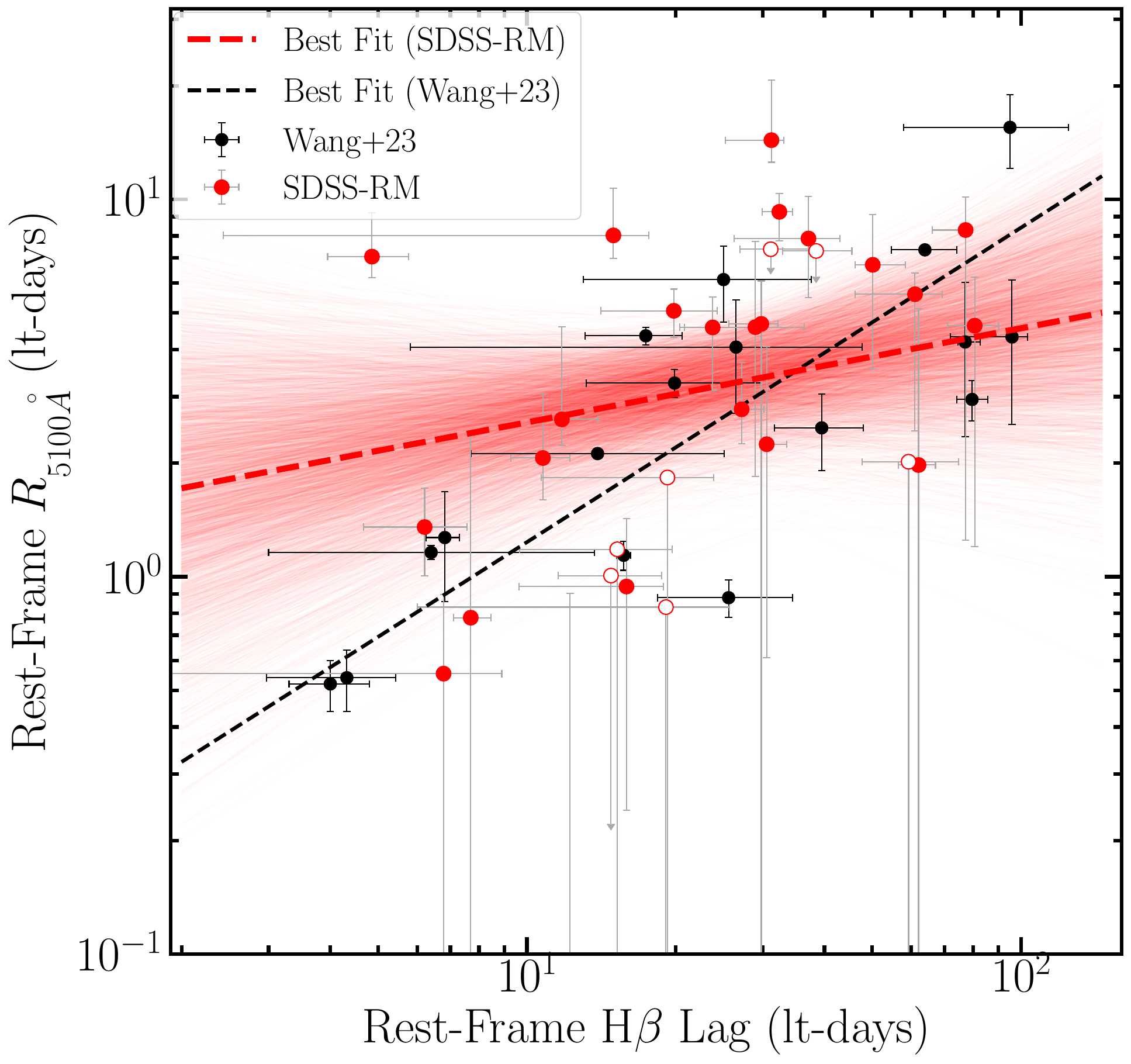}
\caption{Comparison between the 
rest-frame 5100\AA\ continuum and \Hb\ lags
for SDSS-RM quasars (red) and AGN studied by \citet{Wang2023} (black). The black dashed line shows the best-fit relation of \citet{Wang2023}. We perform linear regression on the SDSS-RM measurements, including a survival analysis of 7 lags with upper limits, with the red dashed line indicating the best-fit line and the collection of faint red lines showing the posteriors from the MCMC chain. The best-fit line for SDSS-RM has a slope of $0.25_{-0.23}^{+0.21}$, formally consistent with zero, and $\sigma$ = 0.26 excess scatter. We do not find a significant correlation between the continuum and BLR lags, further indicating that the continuum lags of the SDSS-RM quasars are not dominated by diffuse BLR emission.
}
\label{fig7} 
\end{figure}

Figure~\ref{fig7} presents a comparison of the SDSS-RM targets with the result from \citet{Wang2023}. 
To accommodate upper limits in some of continuum lag measurements as censored data,
we flip the axes between $\tau_{5100}$ and \Hb\ BLR size compared to \citet{Wang2023} and convert their best-fit line to the case where $\tau_{5100}$ is regressed to $\tau_\mathrm{BLR}$. 
Among the 30 SDSS-RM quasars with reliable disk size measurements and \Hb\ BLR measurements, we identify seven quasars with negative continuum lag measurements. We perform a survival analysis using the software package \texttt{PyStan}, treating the seven negative lags as censored data, and we then perform a linear regression that produces a slope of $0.25_{-0.23}^{+0.21}$ (consistent with no correlation)
with $\sigma$ = 0.26 intrinsic scatter. Figure~\ref{fig7} displays the result of our survival analysis and the best-fit line to the total of 30 SDSS-RM quasars.
We find no correlation between continuum lag and \Hb\ lag among the luminous quasars of the SDSS-RM sample, consistent with our previous conclusions that the continuum lags of these quasars are not dominated by diffuse BLR emission.

\subsection{Disk Size and Quasar Properties}
\label{sec:3.3}

To better understand the deviations in the observed disk lags versus those predicted by SS73, we broadly searched for correlations between various quasar properties and logarithmic disk lag ratio ($\log(\tau_{\rm jav}/\tau_{\rm SS73})$). We also tested for correlations between our quasar properties and disk lag offset ($\tau_{\rm jav} - \tau_{\rm SS73}$), as would be appropriate for linearly scaled quasar properties. Our tested quasar properties include optical and X-ray luminosity, black hole mass, Eddington ratio, and various spectral properties associated with ``eigenvector 1'' \citep{Boroson1992,Sulentic2001,Shen2014}, and ionization hardness as inferred from the narrow-line $L\OIII/L(\Hb)$ ratio. We fit each for correlations using linear regression implemented by \texttt{LinMix}, including an intrinsic scatter and with uncertainties sampled by Markov chain Monte Carlo (MCMC). In the case of logarithmic disk lag ratio, we treat the 27 lags with $\tau < 0$ as censored data in the \texttt{LinMix} fits by using their 1$\sigma$ uncertainties as upper limits. The results of our linear fits against log-disk lag ratio are provided in Table~\ref{tab1}.

\begin{table}
\begin{centering}
\hspace{-2 cm}
\begin{tabular}{||c|c|c||} 
\hline
 Quasar Property & $m$ & $\sigma_{\rm scatter}$ \\ [0.5ex] 
\hline\hline
 $\lambda L_{3000}$ & \textbf{-0.38 $\pm$ 0.10} & 0.40 $\pm$ 0.05 \\
\hline
 $L_{\rm X}$ (0.5-2~keV) & -0.36 $\pm$ 0.15 & 0.46 $\pm$ 0.06 \\
\hline
 $M_{\rm BH}$ & \textbf{-0.36 $\pm$ 0.08} & 0.38 $\pm$ 0.05 \\
\hline
 $\lambda_{\rm EDD}$ & -0.02 $\pm$ 0.12 & 0.47 $\pm$ 0.06 \\
\hline
 Fe EW & -0.05 $\pm$ 0.03 & 0.47 $\pm$ 0.06 \\
\hline
Fe/H$\beta$ EW & 0.07 $\pm$ 0.26 & 0.47 $\pm$ 0.06 \\
\hline
H$\beta$ EW & -0.02 $\pm$ 0.19 & 0.48 $\pm$ 0.06 \\
\hline
$L_{\OIII}/L_{\Hb}$ & 0.01 $\pm$ 0.11 & 0.50 $\pm$ 0.07 \\
\hline
He II EW & 0.06 $\pm$ 0.05 & 0.47 $\pm$ 0.06 \\
\hline
\end{tabular}
\caption{Quasar properties that were fit when plotted against $\log(\tau_{\rm jav}/\tau_{\rm obs})$ and the slope of the correlation and excess scatter of the fit including uncertainties found using \texttt{LinMix}. Significant correlations with $m>3\sigma$ are listed in bold.}
\label{tab1}
\end{centering}
\end{table}

\begin{figure*}[ht!]
\centering
    \includegraphics[width=\textwidth]{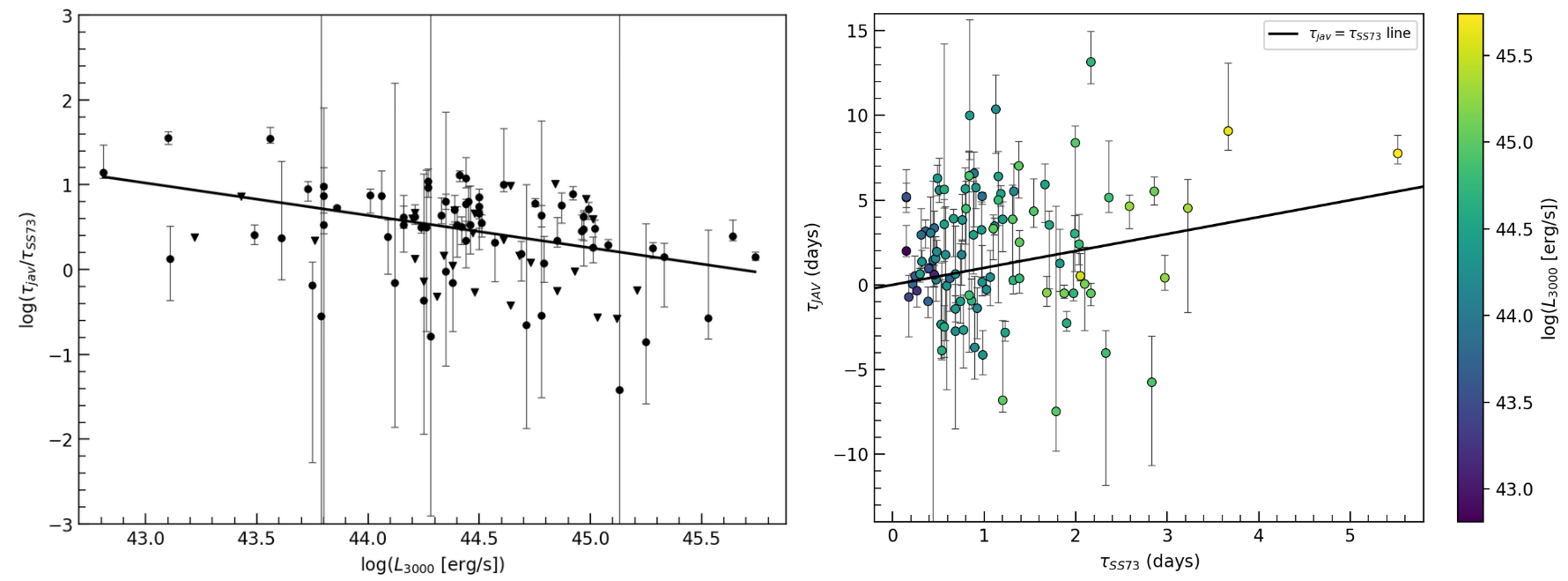}
\caption{The relationship between disk lag and luminosity at 3000 \AA\ ($L_{3000}$). The left plot shows an anti-correlation between log-scaled disk lag ratio ($log(\tau_{\rm jav}/\tau_{\rm SS73})$) and $log(L_{3000})$ using the quasars from \cite{Homayouni2019}, similar to the findings of \cite{Li2021} in their quasar sample. The black triangles with no errorbars represent the upper limit of the quasars which were measured to have $\tau_{\rm jav}<0$. The resulting fit gives a slope of $m = -0.38 \pm 0.10$, and intercept of $b = 18 \pm 0.04$, and intrinsic scatter of $\sigma_{\rm scatter}=0.40 \pm 0.05$. On the right is the measured lag using \texttt{JAVELIN} versus the expected lag from the SS73 model, color-coded by $log(L_{3000})$. The black line represents where observed and model lags would be equal to help interpretation.}\label{fig8}
\end{figure*}

As a result of these linear fits, we identified two correlations $>3\sigma$ significant trends between the tested quasar properties and disk lag ratio, and an additional correlation just shy of the $3\sigma$ level. None of our tested quasar properties correlated with our linear tested quantity disk lag offset however. Figure~\ref{fig8} shows our observed anti-correlation between 3000 \AA\ luminosity and log-disk lag ratio. This correlation was also observed for by \citet{Li2021} in a smaller sample of quasars. We also see an anti-correlation between X-ray luminosity and log-disk lag ratio, implying that disk lag offset may be related to the quasar's bolometric luminosity rather than monochromatic emission. The best-fit linear regression models $\log(\tau_{\rm jav}/\tau_{\rm obs})$ versus 3000 \AA\ and X-ray luminosities find slopes of $m = -0.38 \pm 0.10$, and $m = -0.36 \pm 0.15$ respectively. While the two trends are similar, the 3000 \AA\ correlation is more significant ($>$3$\sigma$ inconsistent with zero). We see a similar correlation with black hole mass, with a slope of $m= -0.36 \pm 0.08$ (Figure~\ref{fig10}).

\begin{figure*}[ht!]
\centering
    \includegraphics[width=\textwidth]{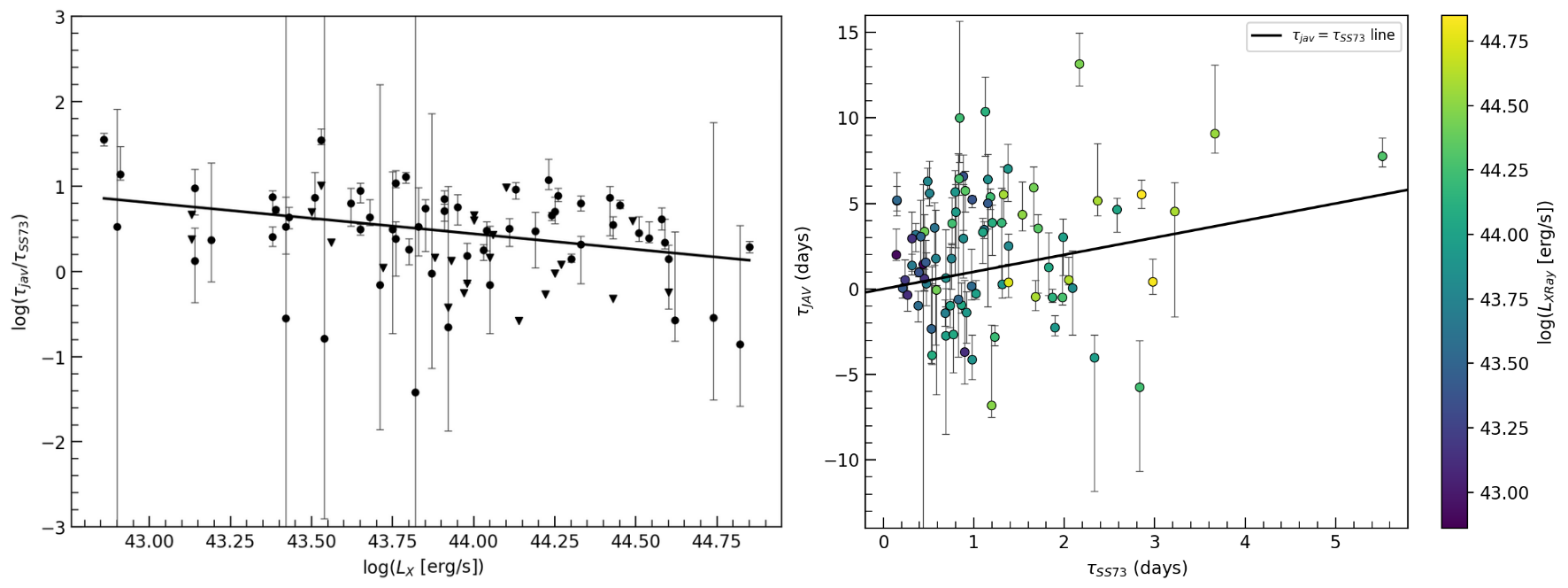}
\caption{The relationship between disk lag and X-ray Luminosity ($L_{\rm X}$) for the 82 of the 95 quasars from \cite{Homayouni2019} that have measured $L_{\rm X}$ from \citet{Liu2020}. The left plot shows an anti-correlation between log-scaled disk lag ratio ($log(\tau_{\rm jav}/\tau_{\rm SS73})$) and $log(L_X)$ similar to the findings of \cite{Li2021} in their quasar sample. The black triangles with no errorbars represent the upper limit of the quasars which were measured to have $\tau_{\rm jav}<0$. The resulting fit gives a slope of $m = -0.36 \pm 0.15$, and intercept of $b = 16 \pm 6$, and intrinsic scatter of $\sigma_{\rm scatter}=0.46 \pm 0.06$. On the right is the measured lag using \texttt{JAVELIN} versus the expected lag from the SS73 model, colored by the $log(L_X)$. The black line represented where the observed and model lags would be equal to help interpretation.}\label{fig9}
\end{figure*}

\begin{figure*}
\centering
    \includegraphics[width=\textwidth]{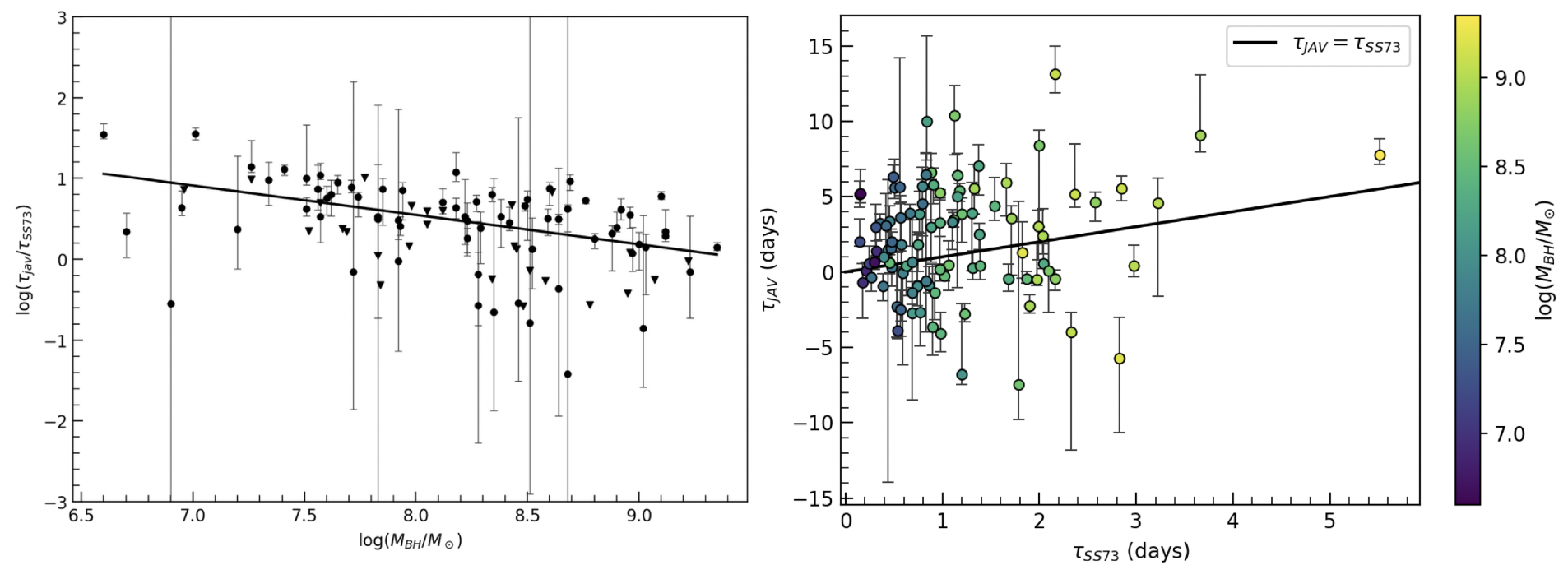}
\caption{The relationship between disk lag and $M_{\rm BH}$. The black triangles with no errorbars represent the upper limit of quasar disk lag ratios which were measured to have $\tau_{\rm jav}<0$. The resulting fit gives a slope of $m = -0.36 \pm 0.08$, and intercept of $b = 3.5 \pm 0.7$, and intrinsic scatter of $\sigma_{\rm scatter}=0.38 \pm 0.05$.}\label{fig10}
\end{figure*}

Each of the anti-correlations found with disk lag ratio exhibit similar slopes of $m \approx -1/3$. The functional form of SS73's predicted disk lag in equation~(\ref{eq:3}) shows $\tau_{\text{SS73}} \propto L_{3000}^{1/3}M_{\rm BH}^{1/3}$. This $m \sim -1/3$ result would thus suggest that when $L_{3000}$ and $M_{\rm BH}$ are fit independently, their anti-correlation is entirely dependent on their exponent in $\tau_{\rm SS73}$. This behavior is highlighted by the right of Figures~\ref{fig8},~\ref{fig9}, and~\ref{fig10} where we show $\tau_{\rm jav}$ as a function of $\tau_{\rm SS73}$ for our 95 quasars. Luminosity and black hole mass do not exhibit any distinguishable behavior with respect to the $\tau_{\rm jav}$ axis. 

While black hole mass and luminosity are connected through Eddington ratio, possibly explaining the similarity in our found anti-correlations, our diverse sample quasar spans $10^{-3}<\lambda_{\rm Edd}<10^{-0.5}$. To better test whether the black hole mass and luminosity anti-correlations are independent, we used a Bayesian maximum likelihood approach to multi-linear regression to fit both simultaneously with disk lag ratio. When using this approach, we rejected all measurements with the $\tau_{\rm jav} < 0$, as the treatment of censored data was not implemented. The multi-linear regression found similar slopes of $m=-0.22 \pm 0.09$ for luminosity, and $m=-0.22 \pm 0.08$ for black hole mass, exhibiting an excess scatter of $\sigma_{\rm excess} = 0.42$. In this case, the multi-linear regression did not favor a singular correlation of black hole mass or luminosity, and thus we conclude that these anti-correlations are likely independent. Our multi-linear regression's shallower slopes may indicate that $\tau_{\rm jav}$ is not completely independent from luminosity and black hole mass, just not as influential as is predicted in $\tau_{\rm SS73}$.


\section{Discussion}
\label{sec:4}

\subsection{Diffuse BLR Contamination}
\label{sec:4.1}

In Section~\ref{sec:3.2}, we tested for the possibility of diffuse BLR emission contaminating the quasar light curves from the Balmer jump in the $i$-band, and the ``pseudo-continuum'' of blended \FeII\ lines in the $g$-band. These particular diffuse emission regions would affect quasars in the redshift range around $0.8<z<1.0$, where our Anderson-Darling test was inconclusive in determining these quasars were drawn from a different population of disk lag offset compared to the whole sample of 95 quasars, shown in Figure~\ref{fig3}. Further investigating the contribution of the Balmer jump and \FeII\ ``pseudo-continuum,'' our measurements of their RMS flux EWs show no correlation with disk lag offset.

Our results exploring diffuse emission differ from those for lower-luminosity local Seyfert 1 AGN, in cases such as NGC 5548 \citep{Fausnaugh2016, Cackett2022} and NGC 4593 \citep{Cackett2018} that find definitive excess lags in the $\sim3650$ \AA\ Balmer jump regime. Such significant diffuse BLR contamination may be more common for these lower luminosity quasars due to the \citet{Baldwin1977} effect, a well known empirical anti-correlation between broad-line strength and luminosity for AGN. Our sample, 
with lags measured between the $g$ and $i$ band corresponding to different rest-frame lags at different redshifts,
should see an increased scatter in disk lag offset due to diffuse contamination affecting either the $g$ or $i$ band. Similar to the behavior of disk lag offset and redshift as shown in Figure~\ref{fig2}, the scatter in disk lag offset is lower for lower luminosity. If we are only considering diffuse Balmer contamination, which for the sample of quasars $z<0.3$ and $L_{3000\AA} < 10^{44}$ erg/s would contaminate the $g$-band, we would expect shorter lags, counter-intuitive to the anti-correlation found in Figure~\ref{fig8}. Our sample shows little evidence of the Baldwin effect increasing diffuse BLR contamination, though 
expanding the comparison to a broader range of
luminosity may probe the Baldwin effect better.

In addition to diffuse BLR emission features, continuum lags can potentially be influenced by more continuous optical emission from the BLR. BLR continuum emission may emit in the form of reflected emission from the accretion disk's continuum emission. Photoionization modeling performed in \citet{Korista2001} suggest this effect should be relatively small, as the effective albedo of broad-line clouds is predicted to be much weaker than diffuse Balmer emission. More recent studies have begun to suggest additional emission across the full UV/optical may originate from the BLR. \citet{Chelouche2019} and \citet{Cackett2022} find that lags need to be described by a combination of disk lags corresponding to the standard accretion disk and the BLR, in which continuous emission from the BLR may originate from high-density material uplifted from the outer accretion disk.
Contributions from BLR continuum emission would increase the measured lag but
cannot explain the large scatter of lags both larger and smaller than SS73 found in our 95 quasar sample. 

\citet{Starkey2023} address the accretion disk size problem by introducing a steep rim to the edge of the accretion disk, as well as potential rippled structures throughout, all irradiated by the AGN's central lamp-post. 
A rim in the outer accretion disk leads to longer lags in a similar way to diffuse BLR continuum emission,
while highly irradiated ripples could satisfy the shorter than expected disk lags found within our sample. 

We also tested our quasar sample for a $R_\mathrm{BLR}-R_{5100\AA}$ correlation, which \citet{Wang2023} find to draw a connection to a $R_{5100\AA}-L$ relation. 
We do not find a significant correlation between the rest-frame continuum lag at 5100\AA\ and the rest-frame \Hb\ lag.
Similar to our other results, this suggests that the continuum lags of SDSS-RM quasars are not dominated by diffuse BLR emission. There may be tight $R_{5100\AA}-L$ relation for lower-luminosity Seyfert 1 AGN, as found by \citet{Wang2023}, but there is not a good correlation for the luminous quasars in the SDSS-RM sample.


\subsection{Anti-Correlations of Disk Lag Ratio with Luminosity and Black Hole Mass}
\label{sec:4.2}

As discussed in Section \ref{sec:3.3}, numerous quasar properties were tested for correlations in the context of linearly-scaled disk lag offset, and log-scaled disk lag ratio. We did not find any significant relations against the linearly-scaled disk lag offset, and further no correlations with accretion rate, probed through various quantities related to eigenvector 1. The anti-correlations found were between log-scaled disk lag ratio vs. optical and X-ray luminosity, as well as black hole mass. These anti-correlations provide interesting implications regarding the accretion disk size problem. Our multi-linear regression fit between disk lag ratio, 3000 \AA\ luminosity, and black hole mass found that neither correlation was dominant over the other. 

In each case, these anti-correlations indicate larger disk lag for fainter and lower mass quasars, with more agreement with SS73 for more luminous and more massive quasars. Our results thus favor models that increase continuum lags for fainter and lower-mass AGN. One possibility for longer lags in less-luminous quasars is the \citet{Baldwin1977} effect, if diffuse BLR emission contributes significantly to the continuum lags.
However, as discussed in Section~\ref{sec:4.1}, diffuse BLR contamination poses the most bias for our lower luminosity and redshift quasars in the $g$-band, which would result in decreased lags with respect to the $i$-band light curves. Diffuse contamination and the Baldwin effect hence should not cause the observed excess lag at lower luminosities found within our quasar sample, as displayed in Figure~\ref{fig8}.
Our sample also shows no evidence for widespread diffuse BLR contamination of the continuum lags, as discussed in Section~\ref{sec:4.1}. That said, our measurements are limited by a single $g$-$i$ lag that probes different rest-frame continuum at different redshifts, and our quasar sample includes quasars of different luminosity at different redshifts. More thoroughly testing for the influence of the Baldwin effect on diffuse BLR contamination requires multi-band continuum RM of quasars spanning a broad range of luminosity.

Another plausible explanation for our observed anti-correlations, as discussed in \cite{Li2021}, is the CHAR model \citep{Sun2020}, which considers magnetohydrodynamic (MHD) heating from a magnetically coupled corona and accretion disk. Such a mechanism would make the observed disk lag dependent on the thermal timescale, which for a steady state disk occurs as $\tau_{\rm TH} \propto L^{0.5}$. In this case, the thermal timescale has the largest deviation from the light crossing time when quasars are less luminous, which would match the anti-correlation shown in Figure~\ref{fig8}.
The observed anti-correlation with black hole mass may have a similar explanation, with smaller disks around smaller black holes showing a more noticeable contribution from $\tau_{\rm TH}$ on top of the light crossing time.

\cite{Kammoun2019} and \cite{Kammoun2021} additionally argue that changing the scale height of the X-ray corona can influence the measured continuum lag.
\citet{Kammoun2021a} predict a positive correlation between measured disk lag and X-ray luminosity, in contrast to our marginal anti-correlation between disk lag and X-ray luminosity.
That said, our observed anti-correlation is marginal ($<$3$\sigma$), 
so definitive conclusions are harder to draw from the fit. 
Additionally, \cite{Kammoun2019} put heavy emphasis on the treatment of ionization in the disk along with corona height.
We tested for correlations with $L\OIII/L(\Hb)$ and \HeII\ EWs as proxies for ionization hardness
but did not find any significant correlations. The lack of observed correlations may be because these quantities are poor proxies for the ionization driven by the central corona and/or because the connection between the corona and the reverberating disk is more complex than can be measured from simple quantities like $L_{\rm X}$.

\section{Summary}

Using the 95 quasars from \cite{Homayouni2019}, we explored how the distribution of disk lag measurements relate with their wide and diverse span of quasar properties across the sample. The results of fitting these quasar properties against the ratio of $\tau_{\rm obs}/\tau_{\rm SS73}$ is summarized in Table~\ref{tab1}. We additionally tested for the possibility of contamination by diffuse BLR emission in the measured continuum lags. The results of this work are the following:

\begin{itemize}

\item Luminosity and black hole mass are anti-correlated with disk lag ratio, as shown in Figures~\ref{fig8}, \ref{fig9}, and \ref{fig10}. The 3000 \AA\ luminosity and black hole mass anti-correlation exhibit a $>3\sigma$ significance, while the X-ray anti-correlation falls just under $3\sigma$, each with slopes $\sim -1/3$. We found no correlation between disk lag ratio with other tested quasar properties associated with ``eigenvector 1'' and ionization hardness.



\item We find no evidence that the continuum lags have widespread contamination from diffuse BLR emission. There is no correlation between the presence of diffuse \FeII\ and Balmer emission in the RMS spectrum with differences in disk lags, and the disk lag offset distribution is consistent for quasars both in and outside the redshift range for which these diffuse BLR features fall in the observed filters. In contrast to \citet{Wang2023}, we do not find a significant correlation between the disk lag and BLR lag.

\end{itemize}

Our results in exploring diffuse contamination and the behavior of various quasar properties with the measured disk lag deviation from SS73 appear to favor the CHAR model \citep{Sun2020}. For our sample, the effects of diffuse BLR contribution to the $g$ and $i$-band photometry has the potential for shorter and longer than expected lags respectively, dependent on the redshift of a given quasar. Our quasar sample reproduces the luminosity anti-correlation with disk lag ratio found in \citet{Li2021}, despite the lower luminosity (lower redshift) quasars being more susceptible to diffuse Balmer contamination in the $g$-band. Diffuse BLR contamination in the bluest lightcurve would result in smaller than expected lags, and as such, the Baldwin effect is a less favorable explanation of this anti-correlation. 

 Lags measured from a single pair of filters have a limited ability to probe the wavelength-dependent contribution of diffuse BLR emission, even among a quasar sample spanning a broad range of redshift. Given diffuse BLR emission has been proven to contribute to continuum light curves and influence lag measurements in low-redshift Seyfert 1 AGN, future multi-band continuum-RM surveys will better determine whether lower luminosity quasars are more prone to diffuse BLR contamination via the Baldwin effect. In addition, frequency resolved lags have the ability roughly probe the reprocessing of particular emitting wavelengths on different time scales. This technique will prove useful in probing BLR photometric contributions and perhaps ripples in the accretion disk for the highest S/N continuum-RM studies.

\acknowledgments

HWS, JRT, MCD, and LBF acknowledge support from NSF grant CAREER-1945546, and with CJG acknowledge support from NSF grants AST-2009539 and AST-2108668. CR acknowledges support from Fondecyt Regular grant 1230345  and ANID BASAL project FB210003. M.L.M.-A. acknowledges financial support from Millenium Nucleus NCN19-058 (TITANs).

Funding for SDSS-III was provided by the Alfred P. Sloan Foundation, the Participating Institutions, the National Science Foundation, and the U.S. Department of Energy Office of Science. The SDSS-III web site is http://www.sdss3.org/.
SDSS-III was managed by the Astrophysical Research Consortium for the Participating Institutions of the SDSS-III Collaboration including the University of Arizona, the Brazilian Participation Group, Brookhaven National Laboratory, Carnegie Mellon University, University of Florida, the French Participation Group, the German Participation Group, Harvard University, the Instituto de Astrofisica de Canarias, the Michigan State/Notre Dame/JINA Participation Group, Johns Hopkins University, Lawrence Berkeley National Laboratory, Max Planck Institute for Astrophysics, Max Planck Institute for Extraterrestrial Physics, New Mexico State University, New York University, Ohio State University, Pennsylvania State University, University of Portsmouth, Princeton University, the Spanish Participation Group, University of Tokyo, University of Utah, Vanderbilt University, University of Virginia, University of Washington, and Yale University.

Funding for the Sloan Digital Sky 
Survey IV has been provided by the 
Alfred P. Sloan Foundation, the U.S. 
Department of Energy Office of 
Science, and the Participating 
Institutions. 

SDSS-IV acknowledges support and 
resources from the Center for High 
Performance Computing  at the 
University of Utah. The SDSS 
website is www.sdss4.org.

SDSS-IV is managed by the 
Astrophysical Research Consortium 
for the Participating Institutions 
of the SDSS Collaboration including 
the Brazilian Participation Group, 
the Carnegie Institution for Science, 
Carnegie Mellon University, Center for 
Astrophysics | Harvard \& 
Smithsonian, the Chilean Participation 
Group, the French Participation Group, 
Instituto de Astrof\'isica de 
Canarias, The Johns Hopkins 
University, Kavli Institute for the 
Physics and Mathematics of the 
Universe (IPMU) / University of 
Tokyo, the Korean Participation Group, 
Lawrence Berkeley National Laboratory, 
Leibniz Institut f\"ur Astrophysik 
Potsdam (AIP),  Max-Planck-Institut 
f\"ur Astronomie (MPIA Heidelberg), 
Max-Planck-Institut f\"ur 
Astrophysik (MPA Garching), 
Max-Planck-Institut f\"ur 
Extraterrestrische Physik (MPE), 
National Astronomical Observatories of 
China, New Mexico State University, 
New York University, University of 
Notre Dame, Observat\'ario 
Nacional / MCTI, The Ohio State 
University, Pennsylvania State 
University, Shanghai 
Astronomical Observatory, United 
Kingdom Participation Group, 
Universidad Nacional Aut\'onoma 
de M\'exico, University of Arizona, 
University of Colorado Boulder, 
University of Oxford, University of 
Portsmouth, University of Utah, 
University of Virginia, University 
of Washington, University of 
Wisconsin, Vanderbilt University, 
and Yale University.

We thank the Bok and CFHT Canadian, Chinese, and French TACs for their support. This research uses data obtained through the Telescope Access Program (TAP), which is funded by the National Astronomical Observatories, Chinese Academy of Sciences, and the Special Fund for Astronomy from the Ministry of Finance in China. This work uses observations obtained with MegaPrime/MegaCam, a joint project of CFHT and CEA/DAPNIA, at the Canada-France-Hawaii Telescope (CFHT) which is operated by the National Research Council (NRC) of Canada, the Institut National des Sciences de l’Univers of the Centre National de la Recherche Scientifique of France, and the University of Hawaii. The authors recognize the cultural importance of the summit of Maunakea to a broad cross section of the Native Hawaiian community. The astronomical community is most fortunate to have the opportunity to conduct observations from this mountain.

\bibliography{refs.bib}{}

\begin{thebibliography}{}
\expandafter\ifx\csname natexlab\endcsname\relax\def\natexlab#1{#1}\fi
\providecommand{\url}[1]{\href{#1}{#1}}
\providecommand{\dodoi}[1]{doi:~\href{http://doi.org/#1}{\nolinkurl{#1}}}
\providecommand{\doeprint}[1]{\href{http://ascl.net/#1}{\nolinkurl{http://ascl.net/#1}}}
\providecommand{\doarXiv}[1]{\href{https://arxiv.org/abs/#1}{\nolinkurl{https://arxiv.org/abs/#1}}}

\bibitem[{{Baldwin}(1977)}]{Baldwin1977}
{Baldwin}, J.~A. 1977, \apj, 214, 679, \dodoi{10.1086/155294}

\bibitem[{{Blandford} \& {McKee}(1982)}]{Blandford1982}
{Blandford}, R.~D., \& {McKee}, C.~F. 1982, \apj, 255, 419,
  \dodoi{10.1086/159843}

\bibitem[{{Boroson} \& {Green}(1992)}]{Boroson1992}
{Boroson}, T.~A., \& {Green}, R.~F. 1992, \apjs, 80, 109,
  \dodoi{10.1086/191661}

\bibitem[{{Cackett} {et~al.}(2021){Cackett}, {Bentz}, \& {Kara}}]{Cackett2021}
{Cackett}, E.~M., {Bentz}, M.~C., \& {Kara}, E. 2021, iScience, 24, 102557,
  \dodoi{10.1016/j.isci.2021.102557}

\bibitem[{{Cackett} {et~al.}(2018){Cackett}, {Chiang}, {McHardy}, {Edelson},
  {Goad}, {Horne}, \& {Korista}}]{Cackett2018}
{Cackett}, E.~M., {Chiang}, C.-Y., {McHardy}, I., {et~al.} 2018, \apj, 857, 53,
  \dodoi{10.3847/1538-4357/aab4f7}

\bibitem[{{Cackett} {et~al.}(2007){Cackett}, {Horne}, \&
  {Winkler}}]{Cackett2007}
{Cackett}, E.~M., {Horne}, K., \& {Winkler}, H. 2007, \mnras, 380, 669,
  \dodoi{10.1111/j.1365-2966.2007.12098.x}

\bibitem[{{Cackett} {et~al.}(2022){Cackett}, {Zoghbi}, \&
  {Ulrich}}]{Cackett2022}
{Cackett}, E.~M., {Zoghbi}, A., \& {Ulrich}, O. 2022, \apj, 925, 29,
  \dodoi{10.3847/1538-4357/ac3913}

\bibitem[{{Chelouche} {et~al.}(2019){Chelouche}, {Pozo Nu{\~n}ez}, \&
  {Kaspi}}]{Chelouche2019}
{Chelouche}, D., {Pozo Nu{\~n}ez}, F., \& {Kaspi}, S. 2019, Nature Astronomy,
  3, 251, \dodoi{10.1038/s41550-018-0659-x}

\bibitem[{{Collier} {et~al.}(1998){Collier}, {Horne}, {Kaspi}, {Netzer},
  {Peterson}, {Wanders}, {Alexander}, {Bertram}, {Comastri}, {Gaskell},
  {Malkov}, {Maoz}, {Mignoli}, {Pogge}, {Pronik}, {Sergeev}, {Snedden},
  {Stirpe}, {Bochkarev}, {Burenkov}, {Shapovalova}, \& {Wagner}}]{Collier1998}
{Collier}, S.~J., {Horne}, K., {Kaspi}, S., {et~al.} 1998, \apj, 500, 162,
  \dodoi{10.1086/305720}

\bibitem[{{Dawson} {et~al.}(2013){Dawson}, {Schlegel}, {Ahn}, {Anderson},
  {Aubourg}, {Bailey}, {Barkhouser}, {Bautista}, {Beifiori}, {Berlind},
  {Bhardwaj}, {Bizyaev}, {Blake}, {Blanton}, {Blomqvist}, {Bolton}, {Borde},
  {Bovy}, {Brandt}, {Brewington}, {Brinkmann}, {Brown}, {Brownstein}, {Bundy},
  {Busca}, {Carithers}, {Carnero}, {Carr}, {Chen}, {Comparat}, {Connolly},
  {Cope}, {Croft}, {Cuesta}, {da Costa}, {Davenport}, {Delubac}, {de Putter},
  {Dhital}, {Ealet}, {Ebelke}, {Eisenstein}, {Escoffier}, {Fan}, {Filiz Ak},
  {Finley}, {Font-Ribera}, {G{\'e}nova-Santos}, {Gunn}, {Guo}, {Haggard},
  {Hall}, {Hamilton}, {Harris}, {Harris}, {Ho}, {Hogg}, {Holder}, {Honscheid},
  {Huehnerhoff}, {Jordan}, {Jordan}, {Kauffmann}, {Kazin}, {Kirkby}, {Klaene},
  {Kneib}, {Le Goff}, {Lee}, {Long}, {Loomis}, {Lundgren}, {Lupton}, {Maia},
  {Makler}, {Malanushenko}, {Malanushenko}, {Mandelbaum}, {Manera}, {Maraston},
  {Margala}, {Masters}, {McBride}, {McDonald}, {McGreer}, {McMahon}, {Mena},
  {Miralda-Escud{\'e}}, {Montero-Dorta}, {Montesano}, {Muna}, {Myers},
  {Naugle}, {Nichol}, {Noterdaeme}, {Nuza}, {Olmstead}, {Oravetz}, {Oravetz},
  {Owen}, {Padmanabhan}, {Palanque-Delabrouille}, {Pan}, {Parejko},
  {P{\^a}ris}, {Percival}, {P{\'e}rez-Fournon}, {P{\'e}rez-R{\`a}fols},
  {Petitjean}, {Pfaffenberger}, {Pforr}, {Pieri}, {Prada}, {Price-Whelan},
  {Raddick}, {Rebolo}, {Rich}, {Richards}, {Rockosi}, {Roe}, {Ross}, {Ross},
  {Rossi}, {Rubi{\~n}o-Martin}, {Samushia}, {S{\'a}nchez}, {Sayres}, {Schmidt},
  {Schneider}, {Sc{\'o}ccola}, {Seo}, {Shelden}, {Sheldon}, {Shen}, {Shu},
  {Slosar}, {Smee}, {Snedden}, {Stauffer}, {Steele}, {Strauss}, {Streblyanska},
  {Suzuki}, {Swanson}, {Tal}, {Tanaka}, {Thomas}, {Tinker}, {Tojeiro},
  {Tremonti}, {Vargas Maga{\~n}a}, {Verde}, {Viel}, {Wake}, {Watson}, {Weaver},
  {Weinberg}, {Weiner}, {West}, {White}, {Wood-Vasey}, {Yeche}, {Zehavi},
  {Zhao}, \& {Zheng}}]{Dawson2013}
{Dawson}, K.~S., {Schlegel}, D.~J., {Ahn}, C.~P., {et~al.} 2013, \aj, 145, 10,
  \dodoi{10.1088/0004-6256/145/1/10}

\bibitem[{Dexter \& Agol(2010)}]{Dexter2010}
Dexter, J., \& Agol, E. 2010, The Astrophysical Journal, 727, L24,
  \dodoi{10.1088/2041-8205/727/1/l24}

\bibitem[{{Di Matteo} {et~al.}(2003){Di Matteo}, Croft, Springel, \&
  Hernquist}]{DiMatteo2003}
{Di Matteo}, T., Croft, R. A.~C., Springel, V., \& Hernquist, L. 2003, The
  Astrophysical Journal, 593, 56, \dodoi{10.1086/376501}

\bibitem[{{Di Matteo} {et~al.}(2005){Di Matteo}, {Springel}, \&
  {Hernquist}}]{DiMatteo2005}
{Di Matteo}, T., {Springel}, V., \& {Hernquist}, L. 2005, \nat, 433, 604,
  \dodoi{10.1038/nature03335}

\bibitem[{{Edelson} {et~al.}(2015){Edelson}, {Gelbord}, {Horne}, {McHardy},
  {Peterson}, {Ar{\'e}valo}, {Breeveld}, {De Rosa}, {Evans}, {Goad}, {Kriss},
  {Brandt}, {Gehrels}, {Grupe}, {Kennea}, {Kochanek}, {Nousek}, {Papadakis},
  {Siegel}, {Starkey}, {Uttley}, {Vaughan}, {Young}, {Barth}, {Bentz},
  {Brewer}, {Crenshaw}, {Dalla Bont{\`a}}, {De Lorenzo-C{\'a}ceres}, {Denney},
  {Dietrich}, {Ely}, {Fausnaugh}, {Grier}, {Hall}, {Kaastra}, {Kelly},
  {Korista}, {Lira}, {Mathur}, {Netzer}, {Pancoast}, {Pei}, {Pogge},
  {Schimoia}, {Treu}, {Vestergaard}, {Villforth}, {Yan}, \& {Zu}}]{Edelson2015}
{Edelson}, R., {Gelbord}, J.~M., {Horne}, K., {et~al.} 2015, \apj, 806, 129,
  \dodoi{10.1088/0004-637X/806/1/129}

\bibitem[{Edelson {et~al.}(2017)Edelson, Gelbord, Cackett, Connolly, Done,
  Fausnaugh, Gardner, Gehrels, Goad, Horne, McHardy, Peterson, Vaughan,
  Vestergaard, Breeveld, Barth, Bentz, Bottorff, Brandt, Crawford, Bont{\`{a}},
  Emmanoulopoulos, Evans, Jaimes, Filippenko, Ferland, Grupe, Joner, Kennea,
  Korista, Krimm, Kriss, Leonard, Mathur, Netzer, Nousek, Page,
  Romero-Colmenero, Siegel, Starkey, Treu, Vogler, Winkler, \&
  Zheng}]{Edelson2017}
Edelson, R., Gelbord, J., Cackett, E., {et~al.} 2017, The Astrophysical
  Journal, 840, 41, \dodoi{10.3847/1538-4357/aa6890}

\bibitem[{Edelson {et~al.}(2019)Edelson, Gelbord, Cackett, Peterson, Horne,
  Barth, Starkey, Bentz, Brandt, Goad, Joner, Korista, Netzer, Page, Uttley,
  Vaughan, Breeveld, Cenko, Done, Evans, Fausnaugh, Ferland, Gonzalez-Buitrago,
  Gropp, Grupe, Kaastra, Kennea, Kriss, Mathur, Mehdipour, Mudd, Nousek,
  Schmidt, Vestergaard, \& Villforth}]{Edelson2019}
---. 2019, The Astrophysical Journal, 870, 123,
  \dodoi{10.3847/1538-4357/aaf3b4}

\bibitem[{{Event Horizon Telescope Collaboration} {et~al.}(2019){Event Horizon
  Telescope Collaboration}, {Akiyama}, {Alberdi}, {Alef}, {Asada}, {Azulay},
  {Baczko}, {Ball}, {Balokovi{\'c}}, {Barrett}, {Bintley}, {Blackburn},
  {Boland}, {Bouman}, {Bower}, {Bremer}, {Brinkerink}, {Brissenden}, {Britzen},
  {Broderick}, {Broguiere}, {Bronzwaer}, {Byun}, {Carlstrom}, {Chael}, {Chan},
  {Chatterjee}, {Chatterjee}, {Chen}, {Chen}, {Cho}, {Christian}, {Conway},
  {Cordes}, {Crew}, {Cui}, {Davelaar}, {De Laurentis}, {Deane}, {Dempsey},
  {Desvignes}, {Dexter}, {Doeleman}, {Eatough}, {Falcke}, {Fish}, {Fomalont},
  {Fraga-Encinas}, {Freeman}, {Friberg}, {Fromm}, {G{\'o}mez}, {Galison},
  {Gammie}, {Garc{\'\i}a}, {Gentaz}, {Georgiev}, {Goddi}, {Gold}, {Gu},
  {Gurwell}, {Hada}, {Hecht}, {Hesper}, {Ho}, {Ho}, {Honma}, {Huang}, {Huang},
  {Hughes}, {Ikeda}, {Inoue}, {Issaoun}, {James}, {Jannuzi}, {Janssen},
  {Jeter}, {Jiang}, {Johnson}, {Jorstad}, {Jung}, {Karami}, {Karuppusamy},
  {Kawashima}, {Keating}, {Kettenis}, {Kim}, {Kim}, {Kim}, {Kino}, {Koay},
  {Koch}, {Koyama}, {Kramer}, {Kramer}, {Krichbaum}, {Kuo}, {Lauer}, {Lee},
  {Li}, {Li}, {Lindqvist}, {Liu}, {Liuzzo}, {Lo}, {Lobanov}, {Loinard},
  {Lonsdale}, {Lu}, {MacDonald}, {Mao}, {Markoff}, {Marrone}, {Marscher},
  {Mart{\'\i}-Vidal}, {Matsushita}, {Matthews}, {Medeiros}, {Menten}, {Mizuno},
  {Mizuno}, {Moran}, {Moriyama}, {Moscibrodzka}, {M{\"u}ller}, {Nagai},
  {Nagar}, {Nakamura}, {Narayan}, {Narayanan}, {Natarajan}, {Neri}, {Ni},
  {Noutsos}, {Okino}, {Olivares}, {Ortiz-Le{\'o}n}, {Oyama}, {{\"O}zel},
  {Palumbo}, {Patel}, {Pen}, {Pesce}, {Pi{\'e}tu}, {Plambeck}, {PopStefanija},
  {Porth}, {Prather}, {Preciado-L{\'o}pez}, {Psaltis}, {Pu}, {Ramakrishnan},
  {Rao}, {Rawlings}, {Raymond}, {Rezzolla}, {Ripperda}, {Roelofs}, {Rogers},
  {Ros}, {Rose}, {Roshanineshat}, {Rottmann}, {Roy}, {Ruszczyk}, {Ryan},
  {Rygl}, {S{\'a}nchez}, {S{\'a}nchez-Arguelles}, {Sasada}, {Savolainen},
  {Schloerb}, {Schuster}, {Shao}, {Shen}, {Small}, {Sohn}, {SooHoo}, {Tazaki},
  {Tiede}, {Tilanus}, {Titus}, {Toma}, {Torne}, {Trent}, {Trippe}, {Tsuda},
  {van Bemmel}, {van Langevelde}, {van Rossum}, {Wagner}, {Wardle},
  {Weintroub}, {Wex}, {Wharton}, {Wielgus}, {Wong}, {Wu}, {Young}, {Young},
  {Younsi}, {Yuan}, {Yuan}, {Zensus}, {Zhao}, {Zhao}, {Zhu}, {Algaba},
  {Allardi}, {Amestica}, {Anczarski}, {Bach}, {Baganoff}, {Beaudoin}, {Benson},
  {Berthold}, {Blanchard}, {Blundell}, {Bustamente}, {Cappallo},
  {Castillo-Dom{\'\i}nguez}, {Chang}, {Chang}, {Chang}, {Chen}, {Chilson},
  {Chuter}, {C{\'o}rdova Rosado}, {Coulson}, {Crawford}, {Crowley}, {David},
  {Derome}, {Dexter}, {Dornbusch}, {Dudevoir}, {Dzib}, {Eckart}, {Eckert},
  {Erickson}, {Everett}, {Faber}, {Farah}, {Fath}, {Folkers}, {Forbes},
  {Freund}, {G{\'o}mez-Ruiz}, {Gale}, {Gao}, {Geertsema}, {Graham}, {Greer},
  {Grosslein}, {Gueth}, {Haggard}, {Halverson}, {Han}, {Han}, {Hao},
  {Hasegawa}, {Henning}, {Hern{\'a}ndez-G{\'o}mez}, {Herrero-Illana},
  {Heyminck}, {Hirota}, {Hoge}, {Huang}, {Impellizzeri}, {Jiang}, {Kamble},
  {Keisler}, {Kimura}, {Kono}, {Kubo}, {Kuroda}, {Lacasse}, {Laing}, {Leitch},
  {Li}, {Lin}, {Liu}, {Liu}, {Lu}, {Marson}, {Martin-Cocher}, {Massingill},
  {Matulonis}, {McColl}, {McWhirter}, {Messias}, {Meyer-Zhao}, {Michalik},
  {Monta{\~n}a}, {Montgomerie}, {Mora-Klein}, {Muders}, {Nadolski}, {Navarro},
  {Neilsen}, {Nguyen}, {Nishioka}, {Norton}, {Nowak}, {Nystrom}, {Ogawa},
  {Oshiro}, {Oyama}, {Parsons}, {Paine}, {Pe{\~n}alver}, {Phillips}, {Poirier},
  {Pradel}, {Primiani}, {Raffin}, {Rahlin}, {Reiland}, {Risacher}, {Ruiz},
  {S{\'a}ez-Mada{\'\i}n}, {Sassella}, {Schellart}, {Shaw}, {Silva}, {Shiokawa},
  {Smith}, {Snow}, {Souccar}, {Sousa}, {Sridharan}, {Srinivasan}, {Stahm},
  {Stark}, {Story}, {Timmer}, {Vertatschitsch}, {Walther}, {Wei}, {Whitehorn},
  {Whitney}, {Woody}, {Wouterloot}, {Wright}, {Yamaguchi}, {Yu}, {Zeballos},
  {Zhang}, \& {Ziurys}}]{EHT2019}
{Event Horizon Telescope Collaboration}, {Akiyama}, K., {Alberdi}, A., {et~al.}
  2019, \apjl, 875, L1, \dodoi{10.3847/2041-8213/ab0ec7}

\bibitem[{{Fausnaugh} {et~al.}(2016){Fausnaugh}, {Denney}, {Barth}, {Bentz},
  {Bottorff}, {Carini}, {Croxall}, {De Rosa}, {Goad}, {Horne}, {Joner},
  {Kaspi}, {Kim}, {Klimanov}, {Kochanek}, {Leonard}, {Netzer}, {Peterson},
  {Schn{\"u}lle}, {Sergeev}, {Vestergaard}, {Zheng}, {Zu}, {Anderson},
  {Ar{\'e}valo}, {Bazhaw}, {Borman}, {Boroson}, {Brandt}, {Breeveld}, {Brewer},
  {Cackett}, {Crenshaw}, {Dalla Bont{\`a}}, {De Lorenzo-C{\'a}ceres},
  {Dietrich}, {Edelson}, {Efimova}, {Ely}, {Evans}, {Filippenko}, {Flatland},
  {Gehrels}, {Geier}, {Gelbord}, {Gonzalez}, {Gorjian}, {Grier}, {Grupe},
  {Hall}, {Hicks}, {Horenstein}, {Hutchison}, {Im}, {Jensen}, {Jones},
  {Kaastra}, {Kelly}, {Kennea}, {Kim}, {Korista}, {Kriss}, {Lee}, {Lira},
  {MacInnis}, {Manne-Nicholas}, {Mathur}, {McHardy}, {Montouri}, {Musso},
  {Nazarov}, {Norris}, {Nousek}, {Okhmat}, {Pancoast}, {Papadakis}, {Parks},
  {Pei}, {Pogge}, {Pott}, {Rafter}, {Rix}, {Saylor}, {Schimoia}, {Siegel},
  {Spencer}, {Starkey}, {Sung}, {Teems}, {Treu}, {Turner}, {Uttley},
  {Villforth}, {Weiss}, {Woo}, {Yan}, \& {Young}}]{Fausnaugh2016}
{Fausnaugh}, M.~M., {Denney}, K.~D., {Barth}, A.~J., {et~al.} 2016, \apj, 821,
  56, \dodoi{10.3847/0004-637X/821/1/56}

\bibitem[{{Fukugita} {et~al.}(1996){Fukugita}, {Ichikawa}, {Gunn}, {Doi},
  {Shimasaku}, \& {Schneider}}]{Fukugita1996}
{Fukugita}, M., {Ichikawa}, T., {Gunn}, J.~E., {et~al.} 1996, \aj, 111, 1748,
  \dodoi{10.1086/117915}

\bibitem[{{Gaskell} \& {Peterson}(1987)}]{Gaskell1987}
{Gaskell}, C.~M., \& {Peterson}, B.~M. 1987, \apjs, 65, 1,
  \dodoi{10.1086/191216}

\bibitem[{{Gaskell} \& {Sparke}(1986)}]{Gaskell1986}
{Gaskell}, C.~M., \& {Sparke}, L.~S. 1986, \apj, 305, 175,
  \dodoi{10.1086/164238}

\bibitem[{{Gravity Collaboration} {et~al.}(2018){Gravity Collaboration},
  {Sturm}, {Dexter}, {Pfuhl}, {Stock}, {Davies}, {Lutz}, {Cl{\'e}net},
  {Eckart}, {Eisenhauer}, {Genzel}, {Gratadour}, {H{\"o}nig}, {Kishimoto},
  {Lacour}, {Millour}, {Netzer}, {Perrin}, {Peterson}, {Petrucci}, {Rouan},
  {Waisberg}, {Woillez}, {Amorim}, {Brandner}, {F{\"o}rster Schreiber},
  {Garcia}, {Gillessen}, {Ott}, {Paumard}, {Perraut}, {Scheithauer},
  {Straubmeier}, {Tacconi}, \& {Widmann}}]{GRAVITY2018}
{Gravity Collaboration}, {Sturm}, E., {Dexter}, J., {et~al.} 2018, \nat, 563,
  657, \dodoi{10.1038/s41586-018-0731-9}

\bibitem[{Grier {et~al.}(2017)Grier, Trump, Shen, Horne, Kinemuchi, McGreer,
  Starkey, Brandt, Hall, Kochanek, Chen, Denney, Greene, Ho, Homayouni, Li,
  Pei, Peterson, Petitjean, Schneider, Sun, AlSayyad, Bizyaev, Brinkmann,
  Brownstein, Bundy, Dawson, Eftekharzadeh, Fernandez-Trincado, Gao,
  Hutchinson, Jia, Jiang, Oravetz, Pan, Paris, Ponder, Peters, Rogerson,
  Simmons, Smith, , \& Wang}]{Grier2017}
Grier, C.~J., Trump, J.~R., Shen, Y., {et~al.} 2017, The Astrophysical Journal,
  851, 21, \dodoi{10.3847/1538-4357/aa98dc}

\bibitem[{Hall {et~al.}(2018)Hall, Sarrouh, \& Horne}]{Hall2018}
Hall, P.~B., Sarrouh, G.~T., \& Horne, K. 2018, The Astrophysical Journal, 854,
  93, \dodoi{10.3847/1538-4357/aaa768}

\bibitem[{{Hern{\'a}ndez Santisteban} {et~al.}(2020){Hern{\'a}ndez
  Santisteban}, {Edelson}, {Horne}, {Gelbord}, {Barth}, {Cackett}, {Goad},
  {Netzer}, {Starkey}, {Uttley}, {Brandt}, {Korista}, {Lohfink}, {Onken},
  {Page}, {Siegel}, {Vestergaard}, {Bisogni}, {Breeveld}, {Cenko}, {Dalla
  Bont{\`a}}, {Evans}, {Ferland}, {Gonzalez-Buitrago}, {Grupe}, {Joner},
  {Kriss}, {LaPorte}, {Mathur}, {Marshall}, {Mehdipour}, {Mudd}, {Peterson},
  {Schmidt}, {Vaughan}, \& {Valenti}}]{Santisteban2020}
{Hern{\'a}ndez Santisteban}, J.~V., {Edelson}, R., {Horne}, K., {et~al.} 2020,
  \mnras, 498, 5399, \dodoi{10.1093/mnras/staa2365}

\bibitem[{Homayouni {et~al.}(2019)Homayouni, Trump, Grier, Shen, Starkey,
  Brandt, Alvarez, Hall, Horne, Kinemuchi, Li, McGreer, Sun, Ho, \&
  Schneider}]{Homayouni2019}
Homayouni, Y., Trump, J.~R., Grier, C.~J., {et~al.} 2019, The Astrophysical
  Journal, 880, 126, \dodoi{10.3847/1538-4357/ab2638}

\bibitem[{Hopkins {et~al.}(2005{\natexlab{a}})Hopkins, Hernquist, Cox, Matteo,
  Martini, Robertson, \& Springel}]{Hopkins2005b}
Hopkins, P.~F., Hernquist, L., Cox, T.~J., {et~al.} 2005{\natexlab{a}}, The
  Astrophysical Journal, 630, 705, \dodoi{10.1086/432438}

\bibitem[{Hopkins {et~al.}(2005{\natexlab{b}})Hopkins, Hernquist, Cox, Matteo,
  Robertson, \& Springel}]{Hopkins2005a}
---. 2005{\natexlab{b}}, The Astrophysical Journal, 630, 716,
  \dodoi{10.1086/432463}

\bibitem[{Jiang {et~al.}(2017)Jiang, Green, Greene, Morganson, Shen, Pancoast,
  MacLeod, Anderson, Brandt, Grier, Rix, Ruan, Protopapas, Scott, Burgett,
  Hodapp, Huber, Kaiser, Kudritzki, Magnier, Metcalfe, Tonry, Wainscoat, \&
  Waters}]{Jiang2017}
Jiang, Y.-F., Green, P.~J., Greene, J.~E., {et~al.} 2017, The Astrophysical
  Journal, 836, 186, \dodoi{10.3847/1538-4357/aa5b91}

\bibitem[{Kammoun {et~al.}(2021{\natexlab{a}})Kammoun, Dov{\v{c}}iak,
  Papadakis, Caballero-Garc{\'{\i}}a, \& Karas}]{Kammoun2021}
Kammoun, E.~S., Dov{\v{c}}iak, M., Papadakis, I.~E., Caballero-Garc{\'{\i}}a,
  M.~D., \& Karas, V. 2021{\natexlab{a}}, The Astrophysical Journal, 907, 20,
  \dodoi{10.3847/1538-4357/abcb93}

\bibitem[{Kammoun {et~al.}(2021{\natexlab{b}})Kammoun, Dov{\v{c}}iak,
  Papadakis, Caballero-Garc{\'{\i}}a, \& Karas}]{Kammoun2021a}
---. 2021{\natexlab{b}}, The Astrophysical Journal, 907, 20,
  \dodoi{10.3847/1538-4357/abcb93}

\bibitem[{{Kammoun} {et~al.}(2019){Kammoun}, {Papadakis}, \&
  {Dov{\v{c}}iak}}]{Kammoun2019}
{Kammoun}, E.~S., {Papadakis}, I.~E., \& {Dov{\v{c}}iak}, M. 2019, \apjl, 879,
  L24, \dodoi{10.3847/2041-8213/ab2a72}

\bibitem[{{Kelly}(2007)}]{Kelly2007}
{Kelly}, B.~C. 2007, \apj, 665, 1489, \dodoi{10.1086/519947}

\bibitem[{{Kelly} {et~al.}(2009){Kelly}, {Bechtold}, \&
  {Siemiginowska}}]{Kelly2009}
{Kelly}, B.~C., {Bechtold}, J., \& {Siemiginowska}, A. 2009, \apj, 698, 895,
  \dodoi{10.1088/0004-637X/698/1/895}

\bibitem[{Kinemuchi {et~al.}(2020)Kinemuchi, Hall, McGreer, Kochanek, Grier,
  Trump, Shen, Brandt, Wood-Vasey, Fan, Peterson, Schneider, Santisteban,
  Horne, Chen, Eftekharzadeh, Guo, Jia, Li, Li, Nie, Ponder, Rogerson, Zhang,
  Zou, Jiang, Ho, Kneib, Petitjean, Palanque-Delabrouille, \&
  Yeche}]{Kinemuchi2020}
Kinemuchi, K., Hall, P.~B., McGreer, I., {et~al.} 2020, The Astrophysical
  Journal Supplement Series, 250, 10, \dodoi{10.3847/1538-4365/aba43f}

\bibitem[{Korista \& Goad(2001)}]{Korista2001}
Korista, K.~T., \& Goad, M.~R. 2001, The Astrophysical Journal, 553, 695,
  \dodoi{10.1086/320964}

\bibitem[{{Kormendy} \& {Ho}(2013)}]{Kormendy2013}
{Kormendy}, J., \& {Ho}, L.~C. 2013, \araa, 51, 511,
  \dodoi{10.1146/annurev-astro-082708-101811}

\bibitem[{{Koz{\l}owski} {et~al.}(2010){Koz{\l}owski}, {Kochanek}, {Udalski},
  {Wyrzykowski}, {Soszy{\'n}ski}, {Szyma{\'n}ski}, {Kubiak}, {Pietrzy{\'n}ski},
  {Szewczyk}, {Ulaczyk}, {Poleski}, \& {OGLE Collaboration}}]{Kozlowski2010}
{Koz{\l}owski}, S., {Kochanek}, C.~S., {Udalski}, A., {et~al.} 2010, \apj, 708,
  927, \dodoi{10.1088/0004-637X/708/2/927}

\bibitem[{{Li} {et~al.}(2021){Li}, {Sun}, {Xu}, {Brandt}, {Trump}, {Yu},
  {Wang}, {Xue}, {Cai}, {Gu}, {Homayouni}, {Liu}, {Wang}, {Zhang}, \&
  {Li}}]{Li2021}
{Li}, T., {Sun}, M., {Xu}, X., {et~al.} 2021, \apjl, 912, L29,
  \dodoi{10.3847/2041-8213/abf9aa}

\bibitem[{{Liu} {et~al.}(2020){Liu}, {Merloni}, {Simm}, {Green}, {Brandt},
  {Schneider}, {Dwelly}, {Salvato}, {Buchner}, {Shen}, {Nandra}, {Georgakakis},
  \& {Ho}}]{Liu2020}
{Liu}, T., {Merloni}, A., {Simm}, T., {et~al.} 2020, \apjs, 250, 32,
  \dodoi{10.3847/1538-4365/abb5b0}

\bibitem[{{Lynden-Bell}(1969)}]{Bell1969}
{Lynden-Bell}, D. 1969, \nat, 223, 690, \dodoi{10.1038/223690a0}

\bibitem[{{MacLeod} {et~al.}(2010){MacLeod}, {Ivezi{\'c}}, {Kochanek},
  {Koz{\l}owski}, {Kelly}, {Bullock}, {Kimball}, {Sesar}, {Westman}, {Brooks},
  {Gibson}, {Becker}, \& {de Vries}}]{MacLeod2010}
{MacLeod}, C.~L., {Ivezi{\'c}}, {\v{Z}}., {Kochanek}, C.~S., {et~al.} 2010,
  \apj, 721, 1014, \dodoi{10.1088/0004-637X/721/2/1014}

\bibitem[{{Magorrian} {et~al.}(1998){Magorrian}, {Tremaine}, {Richstone},
  {Bender}, {Bower}, {Dressler}, {Faber}, {Gebhardt}, {Green}, {Grillmair},
  {Kormendy}, \& {Lauer}}]{Magorrian1998}
{Magorrian}, J., {Tremaine}, S., {Richstone}, D., {et~al.} 1998, \aj, 115,
  2285, \dodoi{10.1086/300353}

\bibitem[{{Markoff} \& {Event Horizon Telescope Collaboration}(2022)}]{EHT2022}
{Markoff}, S., \& {Event Horizon Telescope Collaboration}. 2022, in American
  Astronomical Society Meeting Abstracts, Vol.~54, American Astronomical
  Society Meeting Abstracts, 211.01

\bibitem[{McHardy {et~al.}(2014)McHardy, Cameron, Dwelly, Connolly, Lira,
  Emmanoulopoulos, Gelbord, Breedt, Arevalo, \& Uttley}]{McHardy2014}
McHardy, I.~M., Cameron, D.~T., Dwelly, T., {et~al.} 2014, Monthly Notices of
  the Royal Astronomical Society, 444, 1469, \dodoi{10.1093/mnras/stu1636}

\bibitem[{McHardy {et~al.}(2018)McHardy, Connolly, Horne, Cackett, Gelbord,
  Peterson, Pahari, Gehrels, Goad, Lira, Arevalo, Baldi, Brandt, Breedt, Chand,
  Dewangan, Done, Elvis, Emmanoulopoulos, Fausnaugh, Kaspi, Kochanek, Korista,
  Papadakis, Rao, Uttley, Vestergaard, \& Ward}]{McHardy2018}
McHardy, I.~M., Connolly, S.~D., Horne, K., {et~al.} 2018, Monthly Notices of
  the Royal Astronomical Society, 480, 2881, \dodoi{10.1093/mnras/sty1983}

\bibitem[{Morgan {et~al.}(2010)Morgan, Kochanek, Morgan, \& Falco}]{Morgan2010}
Morgan, C.~W., Kochanek, C.~S., Morgan, N.~D., \& Falco, E.~E. 2010, The
  Astrophysical Journal, 712, 1129, \dodoi{10.1088/0004-637x/712/2/1129}

\bibitem[{Mudd {et~al.}(2018)Mudd, Martini, Zu, Kochanek, Peterson, Kessler,
  Davis, Hoormann, King, Lidman, Sommer, Tucker, Asorey, Hinton, Glazebrook,
  Kuehn, Lewis, Macaulay, Moeller, O'Neill, Zhang, Abbott, Abdalla, Allam,
  Banerji, Benoit-L{\'{e}}vy, Bertin, Brooks, Rosell, Carollo, Kind, Carretero,
  Cunha, D'Andrea, da~Costa, Davis, Desai, Doel, Fosalba,
  Garc{\'{\i}}a-Bellido, Gaztanaga, Gerdes, Gruen, Gruendl, Gschwend,
  Gutierrez, Hartley, Honscheid, James, Kuhlmann, Kuropatkin, Lima, Maia,
  Marshall, McMahon, Menanteau, Miquel, Plazas, Romer, Sanchez, Schindler,
  Schubnell, Smith, Smith, Soares-Santos, Sobreira, Suchyta, Swanson, Tarle,
  Thomas, Tucker, \& and}]{Mudd2018}
Mudd, D., Martini, P., Zu, Y., {et~al.} 2018, The Astrophysical Journal, 862,
  123, \dodoi{10.3847/1538-4357/aac9bb}

\bibitem[{Mummery \& Balbus(2020)}]{Mummery2020}
Mummery, A., \& Balbus, S.~A. 2020, Monthly Notices of the Royal Astronomical
  Society, 492, 5655, \dodoi{10.1093/mnras/staa192}

\bibitem[{{Netzer}(2022)}]{Netzer2022}
{Netzer}, H. 2022, \mnras, 509, 2637, \dodoi{10.1093/mnras/stab3133}

\bibitem[{Peterson(2004{\natexlab{a}})}]{peterson2004}
Peterson, B.~M. 2004{\natexlab{a}}, Proceedings of the International
  Astronomical Union, 2004, 15–20, \dodoi{10.1017/S1743921304001358}

\bibitem[{Peterson(2004{\natexlab{b}})}]{peterson_2004}
---. 2004{\natexlab{b}}, Proceedings of the International Astronomical Union,
  2004, 15–20, \dodoi{10.1017/S1743921304001358}

\bibitem[{{Richards} {et~al.}(2006){Richards}, {Lacy}, {Storrie-Lombardi},
  {Hall}, {Gallagher}, {Hines}, {Fan}, {Papovich}, {Vanden Berk}, {Trammell},
  {Schneider}, {Vestergaard}, {York}, {Jester}, {Anderson}, {Budav{\'a}ri}, \&
  {Szalay}}]{Richards2006}
{Richards}, G.~T., {Lacy}, M., {Storrie-Lombardi}, L.~J., {et~al.} 2006, \apjs,
  166, 470, \dodoi{10.1086/506525}

\bibitem[{Scholz \& Stephens(1987)}]{ADTest}
Scholz, F.~W., \& Stephens, M.~A. 1987, Journal of the American Statistical
  Association, 82, 918.
\newblock \url{http://www.jstor.org/stable/2288805}

\bibitem[{{Sergeev} {et~al.}(2005){Sergeev}, {Doroshenko}, {Golubinskiy},
  {Merkulova}, \& {Sergeeva}}]{Sergeev2005}
{Sergeev}, S.~G., {Doroshenko}, V.~T., {Golubinskiy}, Y.~V., {Merkulova},
  N.~I., \& {Sergeeva}, E.~A. 2005, \apj, 622, 129, \dodoi{10.1086/427820}

\bibitem[{{Shakura} \& {Sunyaev}(1973)}]{SS73}
{Shakura}, N.~I., \& {Sunyaev}, R.~A. 1973, \aap, 500, 33

\bibitem[{{Shappee} {et~al.}(2014){Shappee}, {Prieto}, {Grupe}, {Kochanek},
  {Stanek}, {De Rosa}, {Mathur}, {Zu}, {Peterson}, {Pogge}, {Komossa}, {Im},
  {Jencson}, {Holoien}, {Basu}, {Beacom}, {Szczygie{\l}}, {Brimacombe},
  {Adams}, {Campillay}, {Choi}, {Contreras}, {Dietrich}, {Dubberley},
  {Elphick}, {Foale}, {Giustini}, {Gonzalez}, {Hawkins}, {Howell}, {Hsiao},
  {Koss}, {Leighly}, {Morrell}, {Mudd}, {Mullins}, {Nugent}, {Parrent},
  {Phillips}, {Pojmanski}, {Rosing}, {Ross}, {Sand}, {Terndrup}, {Valenti},
  {Walker}, \& {Yoon}}]{Shappee2014}
{Shappee}, B.~J., {Prieto}, J.~L., {Grupe}, D., {et~al.} 2014, \apj, 788, 48,
  \dodoi{10.1088/0004-637X/788/1/48}

\bibitem[{{Shen} \& {Ho}(2014)}]{Shen2014}
{Shen}, Y., \& {Ho}, L.~C. 2014, \nat, 513, 210, \dodoi{10.1038/nature13712}

\bibitem[{{Shen} {et~al.}(2015){Shen}, {Brandt}, {Dawson}, {Hall}, {McGreer},
  {Anderson}, {Chen}, {Denney}, {Eftekharzadeh}, {Fan}, {Gao}, {Green},
  {Greene}, {Ho}, {Horne}, {Jiang}, {Kelly}, {Kinemuchi}, {Kochanek},
  {P{\^a}ris}, {Peters}, {Peterson}, {Petitjean}, {Ponder}, {Richards},
  {Schneider}, {Seth}, {Smith}, {Strauss}, {Tao}, {Trump}, {Wood-Vasey}, {Zu},
  {Eisenstein}, {Pan}, {Bizyaev}, {Malanushenko}, {Malanushenko}, \&
  {Oravetz}}]{Shen2015}
{Shen}, Y., {Brandt}, W.~N., {Dawson}, K.~S., {et~al.} 2015, \apjs, 216, 4,
  \dodoi{10.1088/0067-0049/216/1/4}

\bibitem[{Shen {et~al.}(2016)Shen, Horne, Grier, Peterson, Denney, Trump, Sun,
  Brandt, Kochanek, Dawson, Green, Greene, Hall, Ho, Jiang, Kinemuchi, McGreer,
  Petitjean, Richards, Schneider, Strauss, Tao, Wood-Vasey, Zu, Pan, Bizyaev,
  Ge, Oravetz, \& Simmons}]{Shen2016}
Shen, Y., Horne, K., Grier, C.~J., {et~al.} 2016, The Astrophysical Journal,
  818, 30, \dodoi{10.3847/0004-637x/818/1/30}

\bibitem[{{Shen} {et~al.}(2019){Shen}, {Hall}, {Horne}, {Zhu}, {McGreer},
  {Simm}, {Trump}, {Kinemuchi}, {Brandt}, {Green}, {Grier}, {Guo}, {Ho},
  {Homayouni}, {Jiang}, {I-Hsiu Li}, {Morganson}, {Petitjean}, {Richards},
  {Schneider}, {Starkey}, {Wang}, {Chambers}, {Kaiser}, {Kudritzki}, {Magnier},
  \& {Waters}}]{Shen2019}
{Shen}, Y., {Hall}, P.~B., {Horne}, K., {et~al.} 2019, \apjs, 241, 34,
  \dodoi{10.3847/1538-4365/ab074f}

\bibitem[{{Soltan}(1982)}]{Soltan1982}
{Soltan}, A. 1982, \mnras, 200, 115, \dodoi{10.1093/mnras/200.1.115}

\bibitem[{Starkey {et~al.}(2015)Starkey, Horne, \& Villforth}]{Starkey2016}
Starkey, D.~A., Horne, K., \& Villforth, C. 2015, Monthly Notices of the Royal
  Astronomical Society, 456, 1960, \dodoi{10.1093/mnras/stv2744}

\bibitem[{{Starkey} {et~al.}(2023){Starkey}, {Huang}, {Horne}, \&
  {Lin}}]{Starkey2023}
{Starkey}, D.~A., {Huang}, J., {Horne}, K., \& {Lin}, D. N.~C. 2023, \mnras,
  519, 2754, \dodoi{10.1093/mnras/stac3579}

\bibitem[{{Sulentic} {et~al.}(2001){Sulentic}, {Calvani}, \&
  {Marziani}}]{Sulentic2001}
{Sulentic}, J.~W., {Calvani}, M., \& {Marziani}, P. 2001, The Messenger, 104,
  25

\bibitem[{Sun {et~al.}(2020)Sun, Xue, Brandt, Gu, Trump, Cai, He, bin Lin, Liu,
  \& Wang}]{Sun2020}
Sun, M., Xue, Y., Brandt, W.~N., {et~al.} 2020, The Astrophysical Journal, 891,
  178, \dodoi{10.3847/1538-4357/ab789e}

\bibitem[{{Tie} \& {Kochanek}(2018)}]{Tie2018}
{Tie}, S.~S., \& {Kochanek}, C.~S. 2018, \mnras, 473, 80,
  \dodoi{10.1093/mnras/stx2348}

\bibitem[{{Vestergaard} \& {Osmer}(2009)}]{Vestergaard2009}
{Vestergaard}, M., \& {Osmer}, P.~S. 2009, \apj, 699, 800,
  \dodoi{10.1088/0004-637X/699/1/800}

\bibitem[{Vincentelli {et~al.}(2022)Vincentelli, McHardy, Santisteban, Cackett,
  Gelbord, Horne, Miller, \& Lobban}]{Vincentelli2022}
Vincentelli, F.~M., McHardy, I., Santisteban, J. V.~H., {et~al.} 2022, Monthly
  Notices of the Royal Astronomical Society: Letters, 512, L33,
  \dodoi{10.1093/mnrasl/slac009}

\bibitem[{Virtanen {et~al.}(2020)Virtanen, Gommers, Oliphant, Haberland, Reddy,
  Cournapeau, Burovski, Peterson, Weckesser, Bright, {van der Walt}, Brett,
  Wilson, Millman, Mayorov, Nelson, Jones, Kern, Larson, Carey, Polat, Feng,
  Moore, {VanderPlas}, Laxalde, Perktold, Cimrman, Henriksen, Quintero, Harris,
  Archibald, Ribeiro, Pedregosa, {van Mulbregt}, \& {SciPy 1.0
  Contributors}}]{SciPy}
Virtanen, P., Gommers, R., Oliphant, T.~E., {et~al.} 2020, Nature Methods, 17,
  261, \dodoi{10.1038/s41592-019-0686-2}

\bibitem[{Wang {et~al.}(2023)Wang, Guo, \& Woo}]{Wang2023}
Wang, S., Guo, H., \& Woo, J.-H. 2023, The Astrophysical Journal Letters, 948,
  L23, \dodoi{10.3847/2041-8213/accf96}

\bibitem[{White \& Peterson(1994)}]{White1994}
White, R.~J., \& Peterson, B.~M. 1994, Publications of the Astronomical Society
  of the Pacific, 106, 879, \dodoi{10.1086/133456}

\bibitem[{{Yu} {et~al.}(2020){Yu}, {Kochanek}, {Peterson}, {Zu}, {Brandt},
  {Cackett}, {Fausnaugh}, \& {McHardy}}]{Yu2020unc}
{Yu}, Z., {Kochanek}, C.~S., {Peterson}, B.~M., {et~al.} 2020, \mnras, 491,
  6045, \dodoi{10.1093/mnras/stz3464}

\bibitem[{Yu {et~al.}(2020)Yu, Martini, Davis, Gruendl, Hoormann, Kochanek,
  Lidman, Mudd, Peterson, Wester, Allam, Annis, Asorey, Avila, Banerji, Bertin,
  Brooks, Buckley-Geer, Calcino, Rosell, Carollo, Kind, Carretero, Cunha,
  D'Andrea, da~Costa, Vicente, Desai, Diehl, Doel, Eifler, Flaugher, Fosalba,
  Frieman, Garc{\'{\i}}a-Bellido, Gaztanaga, Glazebrook, Gruen, Gschwend,
  Gutierrez, Hartley, Hinton, Hollowood, Honscheid, Hoyle, James, Kim, Krause,
  Kuehn, Kuropatkin, Lewis, Lima, Macaulay, Maia, Marshall, Menanteau, Miquel,
  Möller, Plazas, Romer, Sanchez, Scarpine, Schubnell, Serrano, Smith, Smith,
  Soares-Santos, Sobreira, Suchyta, Swann, Swanson, Tarle, Tucker, Tucker, \&
  Vikram}]{Yu2020}
Yu, Z., Martini, P., Davis, T.~M., {et~al.} 2020, The Astrophysical Journal
  Supplement Series, 246, 16, \dodoi{10.3847/1538-4365/ab5e7a}

\bibitem[{{Zu} {et~al.}(2011){Zu}, {Kochanek}, \& {Peterson}}]{Zu2011}
{Zu}, Y., {Kochanek}, C.~S., \& {Peterson}, B.~M. 2011, \apj, 735, 80,
  \dodoi{10.1088/0004-637X/735/2/80}

\end{thebibliography}

\end{document}